# Coexistence and interplay of quantum and classical turbulence in superfluid $^4$He: Decay, velocity decoupling and counterflow energy spectra

S. Babuin[1], V. S. L'vov[2], A. Pomyalov[2], L. Skrbek[3] and E. Varga[3]

[1]*Institute of Physics ASCR, Na Slovance 2, 182 21 Prague, Czech Republic*
[2]*Department of Chemical Physics, The Weizmann Institute of Science, Rehovot 76100 Israel*
[3]*Faculty of Mathematics and Physics, Charles University, Ke Karlovu 3, 121 16 Prague, Czech Republic*
(Dated: June 16, 2016)

We report complementary experimental, numerical and theoretical study of turbulent coflow, counterflow and pure superflow of superfluid $^4$He in a channel, resulting in a physically transparent and relatively simple model of decaying quantum turbulence that accounts for interactions of coexisting quantum and classical components of turbulent superfluid $^4$He. We further offer an analytical theory of the energy spectra of steady-state quantum turbulence in the counterflow and pure superflow, based on algebraic approximation for the energy fluxes over scales. The resulting spectra are not of the classic Kolmogorov form, but strongly suppressed by the mutual friction, leading to the energy dissipation at all scales, enhanced by the counterflow-induced decoupling of the normal- and superfluid velocity fluctuations.

## I. INTRODUCTION

Flows of quantum fluids – such as $^4$He below $T_\lambda \simeq 2.17$ K – displaying superfluidity and the two-fluid behavior, offer a challenging field of fundamental research combining quantum physics and fluid dynamics[1–3]. The phenomenological two-fluid model, suggested by Landau and Tisza, describes dynamics of superfluid $^4$He in terms of interpenetrating normal and superfluid components that have their own densities, $\rho_n(T)$, $\rho_s(T)$, and velocity fields, $\bm{u}_n(\bm{r},t)$, $\bm{u}_s(\bm{r},t)$. In this paper, we consider finite temperature – above about 1 K – where the normal component behaves as a classical fluid with the kinematic viscosity $\nu_n(T)$, while the superfluid component is inviscid, $\nu_s = 0$.

Due to the quantum mechanical restriction, the circulation around the superfluid vortices is quantized in integer values of $\kappa = h/m \simeq 10^{-3}$ cm$^2$/s, where $h$ is the Plank constant and $m$ denotes the mass of $^4$He atom. The singly quantized vortices usually arrange themselves in a tangle that can be characterized by vortex line density (VLD) $\mathcal{L}$, i.e., total length of the quantized vortex line in a unit volume. The dynamical behavior of the tangle constitutes an essential ingredient of *quantum turbulence*, the turbulence occurring in quantum fluids displaying superfluidity. The quantization of circulation in the superfluid component results in appearance of characteristic "quantum" length scale: the mean separation between vortex lines, $\ell = 1/\sqrt{\mathcal{L}}$, which is typically orders of magnitude smaller than the scale $\Delta$ of the largest (energy containing) eddies [4,5].

There is a growing consensus[3,6,7] that the quantization of vortex lines can be neglected at large scales $R \gg \ell$ and that quantum turbulence is similar to classical turbulence if excited similarly, for example, in a steady channel flow by a pressure drop[8–11] or when decaying behind a grid[12,13]. The reason is that the interaction of the normal-fluid component with the quantized vortex tangle leads to a mutual friction force[4,5,14] "which couples together $\bm{u}_n(\bm{r},t)$ and $\bm{u}_s(\bm{r},t)$ so strongly that they move as one fluid" [15]. On the other hand, at small length scales $R \lesssim \ell$, the quantization of vortex lines cannot be neglected and turbulence in superfluids has essentially quantum character.

The pipe and channel flows of viscous fluids belong to the class of most extensively studied classical flows[16,17]. As for pipe and channel flows of superfluids, by combining mechanical and thermal drive, a rich variety of two-fluid turbulent flows with different direction and flow ratio of the two components may be generated, representing a very complex superfluid hydrodynamics system[11,18,19], see Fig. 1. The classical-like mechanical forcing (e.g., by compressing a bellows) results in a *coflow*, the closest analogue to classical viscous channel flow, where both components move, on average, with the same velocity: the mean normal-fluid velocity $\bm{U}_n$ is equal to the mean superfluid velocity $\bm{U}_s$, Fig. 1, right. However, due to quantum-mechanical constraints on the superflow, generated by individual lines, the velocity fields are completely different from the normal-fluid motions at scales $R \lesssim \ell$.

The normal and superfluid components of $^4$He may also be made to flow relative to each other with a non-zero counterflow velocity

$$\bm{U}_{ns} \equiv \bm{U}_n - \bm{U}_s \neq 0 \ . \qquad (1)$$

The *thermal counterflow*, first systematically investigated in pioneering experiments by Vinen[20], is easily generated in a channel with one of its ends sealed and equipped with a heater and open at the other end to a superfluid helium bath (see Fig. 1, left). Here both components move relative to the channel walls. The heat flux is carried away from the heater by the normal fluid alone, and, by conservation of mass, a superfluid current arises in the opposite direction:

$$\rho_n \boldsymbol{U}_n + \rho_s \boldsymbol{U}_s = 0 \,. \qquad (2)$$

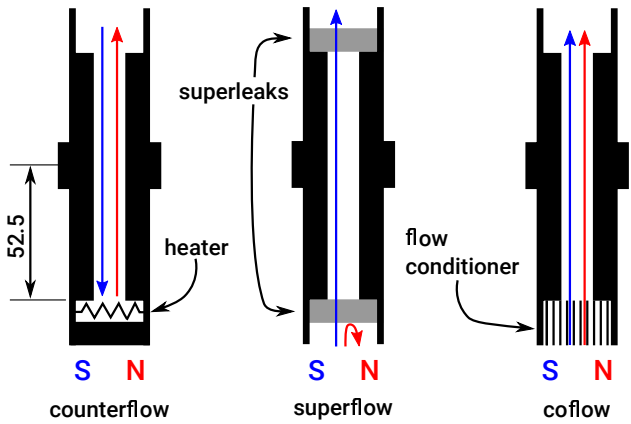

FIG. 1: (color online) Flow channels for the study of counterflow, pure superflow, and coflow. S and N stand for superfluid and normal components. Counterflow is produced thermally by a heater. Superflow and coflow are driven mechanically by a bellows. The turbulence is probed in the middle of the channel by second-sound, excited and detected by mechanical vibration of a porous membrane.

In this way the counterflow velocity $\boldsymbol{U}_{ns}$, proportional to the applied heat flux, is created along the channel, soon accompanied by a tangle of vortex lines. In *pure superflow*, sketched in Fig. 1, middle, superleaks (i.e., filters located at the channel end with sub-micron-sized holes permeable only to the inviscid superfluid component) allow a net flow of the superfluid component in the channel: $\boldsymbol{U}_s \neq 0$, while the normal-fluid component is remaining, on average, at rest: $\boldsymbol{U}_n = 0$. In both cases the fields $\boldsymbol{u}_s(\boldsymbol{r},t)$ and $\boldsymbol{u}_n(\boldsymbol{r},t)$ are expected to be different at all scales. Thermal counterflow and pure superflow therefore represent two special cases of counterflow, characterized by non-zero difference in mean flow velocities of the superfluid and normal components.

In this paper, we report on complementary experimental, analytical and numerical studies of the VLD decay, $\mathcal{L}(t)$, of three categories of turbulent channel flows of superfluid $^4$He, aiming to characterize the quantitative difference between the statistics of turbulence in the coflow on one hand, and the counterflow and superflow on the other. Our study is based on detailed analysis of the VLD decay from three different initial values of $\mathcal{L}_0 \sim 10^4$, $10^5$ and $10^6$ cm$^{-2}$, in all three types of the flows and obtained at different temperatures[21] and represents therefore substantial extension of recent Letter by Gao *et al.*[22], which considered the decaying counterflow. The experimental technique is shortly reviewed in Sec. II A and sketched in Fig. 1.

In Figs. 2 and 3 we present the typical experimental time dependencies of $\mathcal{L}(t)$ decaying by two-three orders of magnitude and analyze them in Sec. II B. These results, partially published in[9,21] are also summarized in Tab. I. We demonstrate that the initial stage of decay in all three types of the flow, including the coflow regime, typically follows a form

$$\mathcal{L}(t) \Rightarrow \mathcal{L}^{\mathrm{Q}}(t) = \frac{b_1 \mathcal{L}_0 |\tau_1|}{t - \tau_1}, \quad \mathcal{L}_0 \equiv \mathcal{L}(t=0)\,, \qquad (3a)$$

with two fitting parameters: the virtual origin time $\tau_1$ and the dimensionless parameter $b_1$. The asymptotics (3a) may be rationalized in the framework of the Vinen evolution equation for $\mathcal{L}(t)$[20] as the decay of random tangle of quantized vortex lines. The energy spectrum of this vortex tangle in the superfluid component (the energy distribution between scales, presented in the $k$-space) $\mathcal{E}_s^{\mathrm{Q}}(k)$ is dominated by the intervortex scales $\ell k \sim 1$.

The late stage of the decay, discussed in Sec. II B, follows a $t^{-3/2}$-asymptotics[13] and may be described by:

$$\mathcal{L}(t) \Rightarrow \mathcal{L}^{\mathrm{cl}}(t) = \frac{b_2 \mathcal{L}_0 |\tau_2|^{3/2}}{(t - \tau_2)^{3/2}}\,, \qquad (3b)$$

with two new fitting parameters: $\tau_2$ and $b_2$. It is commonly believed that the dependence Eq. (3b) is caused by the classical Richardson-Kolmogorov cascade in the superfluid component, with $\mathcal{E}_s^{\mathrm{cl}}(k) \propto k^{-5/3}$ spectrum.

A natural way to rationalize these observations is to assume that the resulting form of the turbulent energy spectrum of the superfluid component $\mathcal{E}_s(k)$ consists of two parts:

$$\mathcal{E}_s(k) = \mathcal{E}_s^{\mathrm{Q}}(k) + \mathcal{E}_s^{\mathrm{cl}}(k)\,, \qquad (4)$$

(i) the classical region spanning large scales from the integral length-scale $\Delta$ down to the intervortex distance $\ell$, where it is followed by (ii) a quantum contribution $\mathcal{E}_s^{\mathrm{Q}}(k)$, corresponding to the random tangle of quantized vortex lines, having a form of a peak. Qualitatively, this energy spectrum is sketched in Fig. 4(a).

Section III is devoted to the theoretical discussion on the stationary energy spectra $\mathcal{E}_s(k)$ in the superfluid component of $^4$He quantum turbulence. The main aim of this Section is to clarify the striking difference between $\mathcal{E}_s(k)$ in the coflow from that in the counterflow and superflow. In Sec. III A, we explain why the energy spectrum of channel turbulence in coflow may include, besides the classical Kolmogorov spectrum $\mathcal{E}^{\mathrm{K41}}(k) \propto k^{-5/3}$, the quantum peak centered at wave-numbers about $1/\ell$. Additionally, we argue that this peak has large $k$ asymptotics $\propto 1/k$, originating from the superflow near the cores of quantized vortices, while for smaller $k$ it borders the classical spectrum $\propto k^2$, caused by the thermodynamical equilibrium with the energy equipartition between degrees of freedom, see Fig. 4(a). In Sec. III B we argue that for stationary regimes of the counterflow and pure superflow, the energy spectrum of the superfluid component has a qualitatively different form from that in the coflow, as sketched in Fig. 4(b). The physical reason for that is the decoupling of the normal and superfluid turbulent velocity fluctuations, caused by the non-zero value of the counterflow velocity $\boldsymbol{U}_{\rm ns}$ (1) in the counterflow and superflow, while in the coflow $\boldsymbol{U}_{\rm ns} = 0$. This position is strongly supported by recent analytical research[23], reviewed in Sec. III B 1. A short discussion of a possible form of the energy spectrum, also following Ref.[23], is given in Sec. III B 2.

At this point, in Sec IV we are armed to formulate, to discuss and to compare with experiments two models of decaying quantum turbulence. First of all, based on explained above forms of the energy spectra for coflow, Fig. 4(a), we suggest in Sec. IV A a "basic model" of the VLD decay, Eq. (25), in which the quantum decay term (21a) and the classical energy source term (24e) are present from the very beginning of the decay, $t = 0$. This model reproduces both quantum and classical asymptotics, given by Eqs. (3), in agreement with the observations in coflow, shown in Figs. 2 and 3. Based on the basic model, in Sec. IV B we present a more detailed (but still preliminary) discussion of the underlying physics, hidden in its fitting parameters. Although the model explains small- and large-time asymptotics of the $\mathcal{L}(t)$ dependence, it fails to describe the crossover regime between them, even for the coflow.

The most striking disagreement between the simple "basic" model and observations, seen in Fig. 3, is a "bump" (non-monotonic behavior) in the $\mathcal{L}(t)$ dependence, in the counterflow and superflow cases. This behavior may be naturally explained by the delay in the delivery of the energy flux from the classical to the quantum part of the spectra (as it has been shown in a slightly different manner in the recent Letter by Gao *et al.*[22] for the special case of decaying thermal counterflow) required for evolution from the more localized in $k$ spectrum, sketched in Fig. 4(b), toward the K41-spectrum, shown in Fig. 4(a) and (c). This delay is analyzed numerically in Sec. IV C and is accounted for in the "improved model" of the VLD decay, formulated in Sec. IV D. In this Section we show that $\mathcal{L}(t)$-dependence, following from the improved model, allows us to rationalize the main experimental observations including the bump in the $\mathcal{L}(t)$-dependence at the crossover times.

Following analysis of our experimental findings, we suggest in Sec. V an analytical theory of the energy spectra for the steady-state quantum turbulence in the counterflow and pure superflow, based on the algebraic approximation for the energy fluxes over scales. The resulting spectra are strongly suppressed by the mutual friction, leading to the energy dissipation at all scales, enhanced by the counterflow-induced decoupling of the normal- and superfluid velocity fluctuations.

In the final Sec. VI we summarize our main results and discuss issues that remain out of the scope of this research. In particular, we stress that our simple analytical theory of the steady-state energy spectra of quantum turbulence in the counterflow adopts some uncontrolled approximations and simplifications, widely used in the studies of classical hydrodynamic turbulence. In the further studies of quantum turbulence these assumptions have to be either better justified or relaxed. Nevertheless, we consider our findings as a natural and perhaps even required step in this direction.

## II. DECAY OF THE VORTEX TANGLE IN COFLOW, COUNTERFLOW AND SUPERFLOW

### A. Experiment

The superflow and the coflow of superfluid $^4$He are both mechanically forced by a low temperature bellows through a square cross-section brass channel, illustrated in Fig. 1. The same channel can be configured to attain superflow, coflow and thermal counterflow. For superflow, sintered silver filters – superleaks – block the viscous normal component. Two vertical brass flow channels have been used, both with the test section 105 mm in length and has an internal square cross-section of side 7 mm and 10 mm, therefore with a factor 2 change in cross-sectional area. For coflow, the superleaks are removed and the lower one is replaced by a flow conditioner made from a dense pack of 10 mm long capillaries of 1 mm diameter, intended to cut larger scale turbulent eddies. The counterflow is prepared by closing one end of the channel with a resistive wire heater and leaving the other end open. A full description of the mechanically driven superflow apparatus and the measurement technique is given in Ref.[24]. Counterflow is studied as in previous Prague experiments[25,26].

Turbulence is detected by measuring the extra attenuation of second sound caused by the scattering of normal-component thermal excitations by the vortex lines. Second sound is generated and detected by a pair of vibrating porous membranes located in the walls of the channel at its mid-point; the second sound travels across the channel, which acts as a resonator. The time dependent attenuated amplitude of second sound at resonance $a(t)$ can be related to the instantaneous VLD $\mathcal{L}(t)$ (assuming random and not extremely dense tangle[24]) through the equation:

$$\mathcal{L}(t) = \frac{6\pi \Delta f_0}{B\kappa} \left[ \frac{a_0}{a(t)} - 1 \right], \qquad (5)$$

where $a_0$ and $\Delta f_0$ are the amplitude and full width at half maximum of the second sound amplitude resonant curve for quiescent helium, and $B$ is the mutual friction coefficient of order unity, tabulated in Ref.[27]. The attenuation of second sound measures the length of vortex line per unit volume weighted by a factor $\sin^2 \theta$, where $\theta$ is the angle between any element of vortex line and the direction of propagation of the second sound.

| 1 | 2 | 3 | 4 | 5 | 6 | 7 | 8 | 9 | 10 | 11 | 12 | 13 | 14 | 15 | 16 | 17 | 18 | 19 |
|---|---|---|---|---|---|---|---|---|---|---|---|---|---|---|---|---|---|---|
| # | type of the flow | $T$, K | $U_s$, cm/s | $U_n$, cm/s | $U_{ns}$ cm/s | $\mathcal{L}_0 \cdot 10^{-4}$ cm$^{-2}$ | Re$_\tau$ — | $b_1$ — | $\tau_1$ s | $b_2$ — | $\tau_2$ s | $\mathcal{L}_0/\mathcal{L}_1$ — | $\mathcal{L}_1/\mathcal{L}_2$ — | $t_1$ s | $t_2$ s | $d_1$ — | $d_2$ — | $\mathcal{L}_0/\mathcal{L}_2$ — |
| 1 |  |  | 0.69 | 0.69 | 0 | 0.86 | 139 | -1.27 | -1.26 | - 2.22 | -0.08 | 2.3 | 3.9 | 2.0 | 14 | 0.54 | -0.0035 | 8.97 |
| 2 |  | 1.35 | 4.99 | 4.99 | 0 | 13.3 | 797 | -1.19 | -0.26 | - 0.22 | -0.05 | 5.9 | 4.3 | 1.4 | 5.3 | 1.73 | -0.019 | 25.4 |
| 3 | co- |  | 22.4 | 22.4 | 0 | 106 | 3092 | -1.15 | -0.11 | - 0.001 | 1.15 | 12.5 | 4.7 | 1.3 | 3.6 | 5.36 | 0.0049 | 58.8 |
| 4 | flow, |  | – | – | 0 | 6.0 | 843 | -1.04 | -0.8 | -0.03 | -1.0 | – | – | – | 1.7 | 3.05 | -0.0092 | 4.16 |
| 5 | 7mm | 1.45 | 4.99 | 4.99 | 0 | 20.0 | 3276 | -1.09 | -0.28 | - 0.05 | -0.2 | 5.6 | 3.6 | 1.6 | 4.5 | 3.70 | -0.0039 | 20.2 |
| 6 |  |  | 22.4 | 22.4 | 0 | 80.0 | – | - 1.25 | -0.13 | 0.006 | 0.5 | 7.7 | 13 | 1.2 | 5.0 | 7.92 | 0.0046 | 100 |
| 7 | counter- |  | −0.09 | 0.95 | 1.05 | 0.96 | 0.96 | - 1.02 | -0.41 | 0.03 | 6.0 | 2.0 | 7.0 | 1.5 | 2.7 | 0.24 | 0.0547 | 14.0 |
| 8 | flow, | 1.45 | −0.29 | 2.89 | 3.18 | 11.2 | 775 | - 0.90 | -0.04 | 0.18 | 0.1 | 6.3 | 2.9 | 0.7 | 8.1 | 0.25 | 0.0070 | 18.3 |
| 9 | 10mm |  | −0.77 | 7.77 | 8.54 | 92.5 | 1888 | -0.61 | -0.013 | -0.25 | -0.01 | 13.2 | 38 | 0.1 | 3.1 | 0.45 | -0.0019 | 502 |
| 10 | super- |  | 0.96 | 0 | 0.96 | 1.21 | 176 | - 1.40 | -0.18 | - 0.01 | -7 | 6.3 | 2.9 | 0.4 | 32 | 0.18 | -0.0340 | 18.3 |
| 11 | flow, | 1.45 | 2.67 | 0 | 2,67 | 10.4 | 467 | 0.49 | 0.07 | -0.003 | -5.8 | 17.2 | 3.9 | 0.2 | 6.0 | 0.22 | -0.0516 | 67.1 |
| 12 | 10mm |  | 7.41 | 0 | 7.41 | 113 | 1246 | 0.25 | 0.04 | -0.015 | -0.15 | 9.1 | 5.5 | 0.1 | 12 | 0.68 | -0.0065 | 50.1 |

TABLE I: Column #2 shows types of the flow and channel width of chosen twelve sets of experiments, numbered in first column from 1 to 12). Next 17 columns display: #3 – Temperature $T$ in K; #4, #5 and #6 – superfluid, normal fluid and counterflow velocities $U_s$, $U_n$ and $U_{ns}$ in cm/s; #7 – initial VLD $\mathcal{L}_0$ in cm$^{-2}$; #8 – Reynolds number Re$_\tau$, estimated via counterflow velocity $U_{ns}$ and normal-fluid kinematic viscosity $\nu_n$; ##8-12: parameters of the fits (3a) for the Vinen's ($\propto t^{-1}$) decay and for the hydrodynamic ($\propto t^{-3/2}$) decay(3b), namely #9 and #11 – dimensionless $b_1$ and $b_2$ and #10 and # 12 – time-origins $\tau_1$ and $\tau_2$ in seconds; #13 – ratio of the initial VLD $\mathcal{L}_0$ to $\mathcal{L}_1$, lowest value of $\mathcal{L}(t)$ in the $t^{-1}$-fit at the time $t = t_1$ (shown in #15); #14 – ratio of the VLD $\mathcal{L}_1$ to $\mathcal{L}_2$, the initial (largest) value of $\mathcal{L}(t)$ in the $t^{-3/2}$-fit at the time $t = t_2$(shown in #16); #17 and #18 – model parameters $d_1$ and $d_2$, related to $b_1$ and $b_2$ by Eqs. (22) and (26b) (with $\alpha = 0.06$ for $T = 1.45$ K and $\alpha = 0.04$ for $T = 1.35$ K). #19 – ratio of the initial VLD $\mathcal{L}_0$ to $\mathcal{L}_2$.

We have performed mechanically-driven coflow, superflow and thermally driven counterflow decay measurements, in the two channels of widths $2\Delta = 7$ mm and 10 mm, in the temperature range between 1.25 K to 2.10 K, and for different values of velocities chosen such to produce initial steady-state VLD $\mathcal{L}_0$, spaced almost exactly one decade apart: $10^4$ cm$^{-2}$, $10^5$ cm$^{-2}$ and $10^6$ cm$^{-2}$. For every combination of temperature and starting line density we have measured typically 150 individual decays, under nominally identical experimental conditions and we have ensemble-averaged these samples, allowing us to resolve up to 4 orders of magnitude of decay on $\mathcal{L}(t)$.

### B. Experimental data and their preliminary analysis

In this paper, we restrict ourselves by discussing 12 typical data sets of experiments, characterized in Tab. I, which are chosen to illustrate the basic ideas of this research. For $T = 1.45$ K we discuss three sets with $\mathcal{L}_0 \sim 10^4$, $10^5$ and $10^6$ cm$^{-2}$ for coflow (sets # 3, 4 and 5), counterflow (sets # 7, 8 and 9) and superflow (sets # 10, 11 and 12). In addition, we analyze the coflow sets for lower $T = 1.35$ K with the same $\mathcal{L}_0 \sim 10^4$, $10^5$ and $10^6$ cm$^{-2}$ (sets # 1, 2 and 3).

The experiments were conducted in the channels of two different widths. For coflow we have more representative data taken from experiments with 7 mm channel, while for superflow and counterflow more representative data were obtained with 10 mm channel. We found no significant differences between 7 mm and 10 mm channel results for a given flow type and therefore compare below the most representative results regardless of the channel width.

#### 1. Coflow

We begin to analyze the time evolution of $\mathcal{L}(t)$ from the decaying coflow turbulence, which, in some sense, is the simplest case. The VLD $\mathcal{L}(t)$ is decaying monotonically, as illustrated in Fig. 2 for both temperatures $T = 1.35$ K [shown in Figs. 2(a) and 2(b)] and $T = 1.45$ K [shown in Figs. 2(c) and 2(d)] and for all three initial values of $\mathcal{L}_0$: $10^4$ cm$^{-2}$ (blue lines), $10^5$ cm$^{-2}$ (green lines) and $10^6$ cm$^{-2}$ (red lines).

There is no qualitative difference between two presented temperatures, except that for higher $T = 1.45$ K the VLD $\mathcal{L}(t)$ is decaying slightly faster. Thus there is no reason to analyze these cases separately.

Our first step is an analysis of the initial stage of the coflow turbulence decay shown in Fig. 2(b) and Fig. 2(d) (for $T = 1.35$ K and $T = 1.45$ K) fitted by $t^{-1}$-law (3a), as shown by dashed lines. The fitting parameters $b_1$ and $\tau_1$, given in Tab. I, will be discussed later. For now, we notice that the negative virtual origin time $\tau_1$ increases (and becomes closer to zero) with increasing $\mathcal{L}_0$, as expected from Eq. (3a).

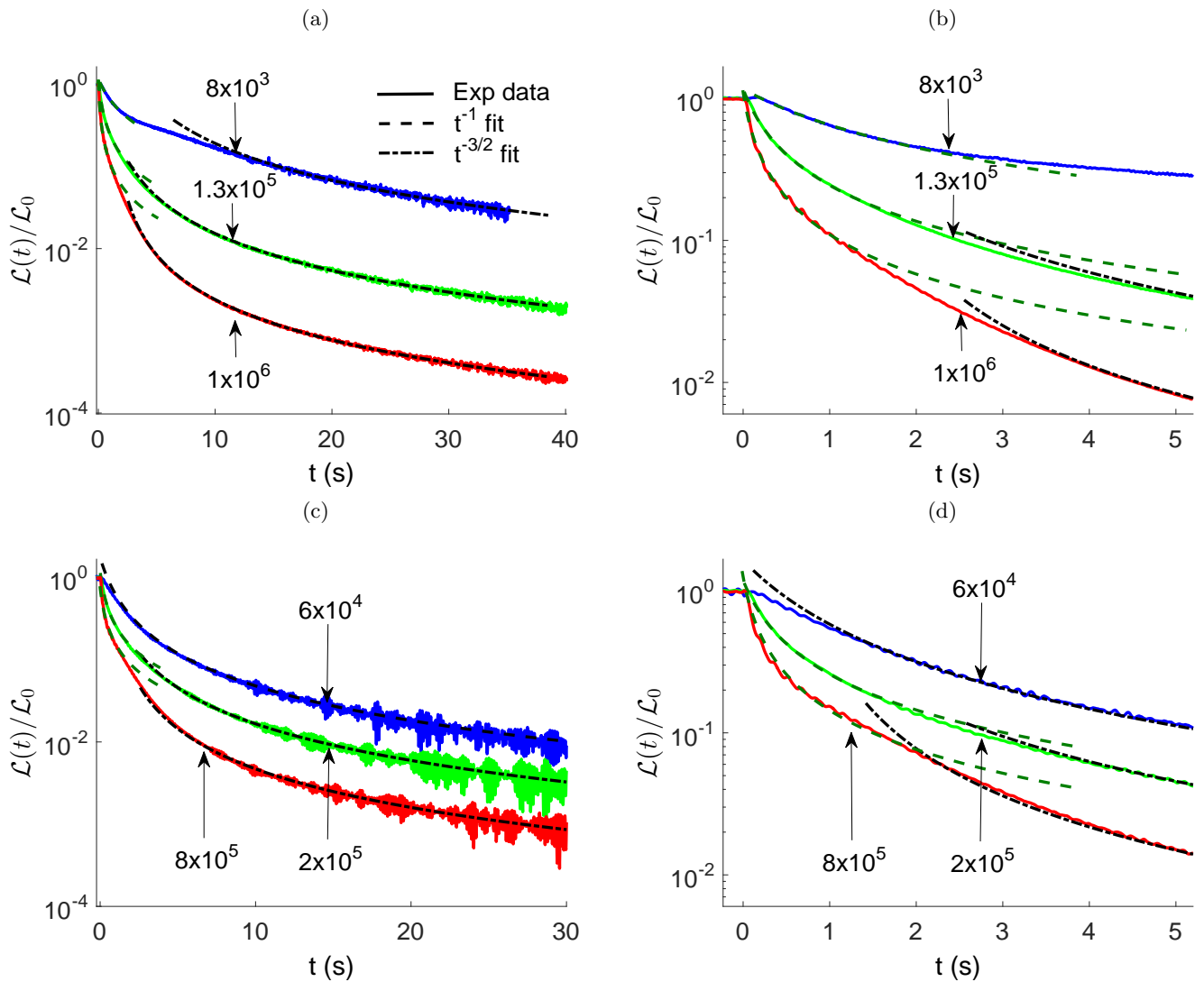

FIG. 2: (color online) Experimental data of the VLD decay $\mathcal{L}(t)/\mathcal{L}_0$ in the coflow (in the 7 mm channel), normalized by initial VLD $\mathcal{L}_0$. The lines correspond (from top to bottom) for $\mathcal{L}_0 \simeq 10^4$ (blue lines), $10^5$ (green lines) and $10^6$ (red lines). The explicit values of $\mathcal{L}_0$ are shown in figures. Data for $T = 1.35$ K are shown in panels (a) and (b), and for $T = 1.45$ K – in panels (c) and (d). The panels (b) and (d) emphasise the details of the short-time behavior. Quantum $t^{-1}$-fits, Eq. (3a), are shown by dashed dark green lines, while the classical $t^{-3/2}$-fits, Eq. (3b) – by black dot-dashed lines.

As seen in Figs. 2(b) and 2(d), the "quantum" $t^{-1}$-fit (3a) agrees with the experimental data over the time interval $0 \lesssim t \lesssim t_1$ about $1 \div 2$s, when $\mathcal{L}(t)$ decays from $\mathcal{L}_0$ to $\mathcal{L}(t_1) \equiv \mathcal{L}_1$. Ratios of the initial and final values of the VLD in the quantum decay, $\mathcal{L}_0/\mathcal{L}_1$ and final time $t_1$, are given in Tab. I. One sees that for the largest $\mathcal{L}_0 \sim 10^6$ cm$^{-2}$ the ratio $\mathcal{L}_0/\mathcal{L}_1$ reaches one order of magnitude. On the other hand, sometimes (e.g. set # 4 with $T = 1.45$ K and smallest $\mathcal{L}_0 \simeq 6 \cdot 10^4$), the quantum regime of the coflow decay does not show up. The consequences of these important observations will be discussed in Sec. IV.

The second step is the analysis of the later stage of the coflow turbulence decay shown by dash-dotted lines for $T = 1.35$ K and $T = 1.45$ K in Figs. 2(a), 2(c) and fitted from time $t_2$ by $t^{-3/2}$-law (3b). In Tab. I we present the fitting parameters $b_2$ and $\tau_2$, the starting fit-time $t_2$ together with the ratio of the final value $\mathcal{L}_1 = \mathcal{L}(t_1)$ in the quantum fit to the initial value $\mathcal{L}_2 \equiv \mathcal{L}(t_2)$ in the classical fit.

The intermediate regime between the quantum and the classical ones (lasting from $t_1$ to $t_2$ and during which $\mathcal{L}(t)$ decays from $\mathcal{L}_1$ to $\mathcal{L}_2$) is the subject of a separate discussion in the Sec. IV.

Returning to the classical stage of the decay, notice that similar to the case of quantum decay, the origin time $\tau_2$ for $\mathcal{L}_0 \sim 10^5$ is larger (closer to zero) than that for smaller $\mathcal{L}_0 \sim 10^4$, as expected from Eq. (3b). However, for larger $\mathcal{L}_0 \sim 10^6$, the time $\tau_2$ becomes positive, which formally contradicts Eq. (3b). This can be explained assuming that there are some transient processes between quantum and classical stages of the decay that become significant for large initial values of $\mathcal{L}_0$. A possible physical reason for this will be suggested in Sec. IV, after discussion of the steady-state energy spectra of quantum turbulence in the coflow.

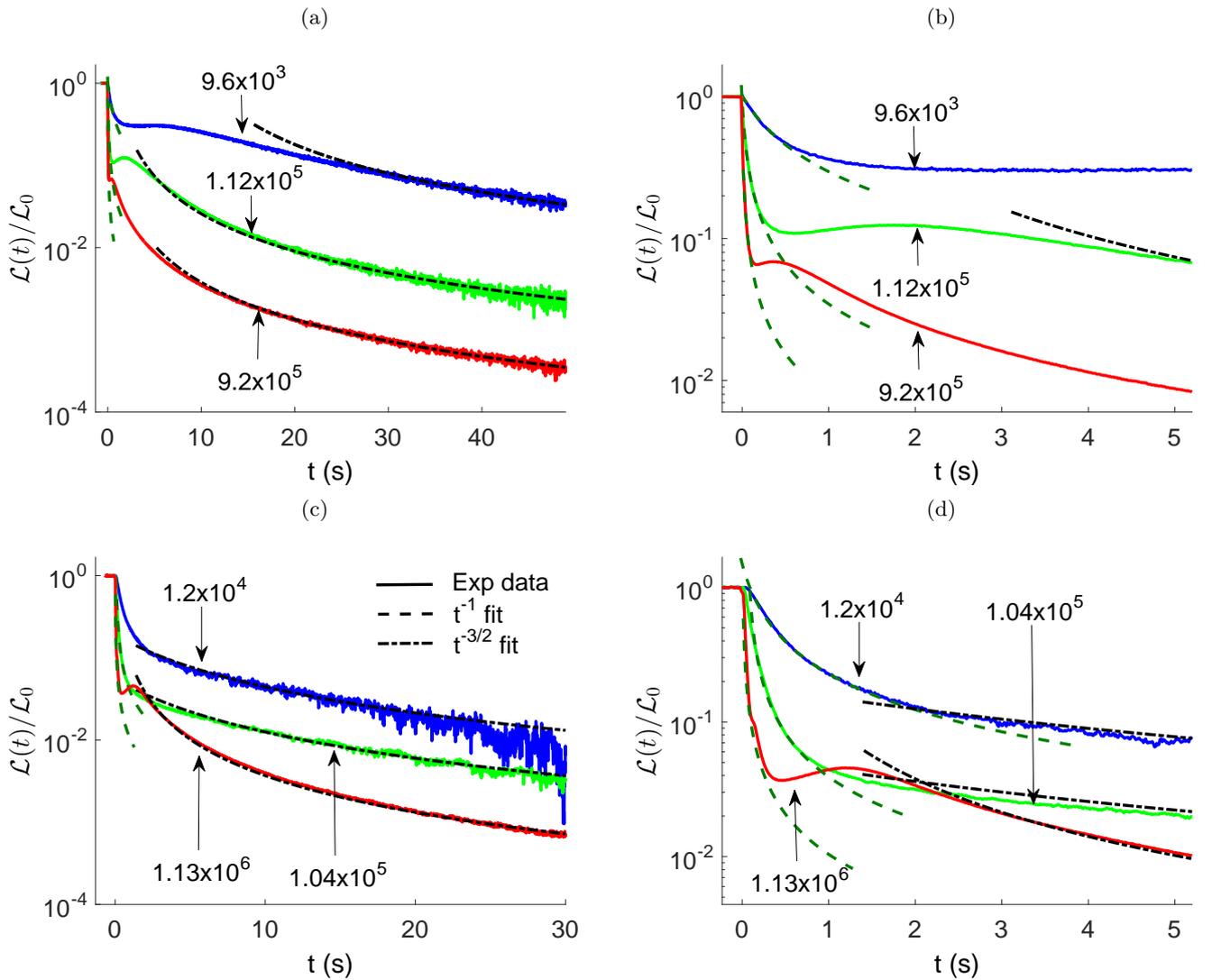

FIG. 3: (color online) Experimental data for the VLD decay $\mathcal{L}(t)/\mathcal{L}_0$ at $T=1.45$ K, normalized by initial VLD $\mathcal{L}_0$. The results for counterflow are shown in panels(a) and (b), for superflow - in panels(c) and (d). In all panels the lines correspond to (from top to bottom) $\mathcal{L}_0 \simeq 10^4$ (blue lines), $10^5$ (green lines) and $10^6$ (red lines). The explicit values of $\mathcal{L}_0$ are shown in the figure. As in Fig. 2, quantum $t^{-1}$-fits, Eq. (3a), are shown by dashed dark green lines, while the classical $t^{-3/2}$-fits, Eq. (3b) – by black dot-dashed lines. The panels (b) and (d) show the initial stages of the decay.

Comparing the rates of the decay at early (quantum) and late (classical) stages in Figs. 2, one sees that classical decay is slower than the quantum one. The physical reason for that is simple: in the quantum regime the energy source in the decaying vortex tangle is relatively small with respect of its dissipation rate and may be neglected. Later, when the dissipation rate, proportional to $\mathcal{L}^2(t)$, becomes smaller, the Richardson-Kolmogorov energy cascade toward small scales, serving as an additional energy source for the vortex tangle, should be taken into account.

From the formal point of view, it looks contradictory that the $t^{-3/2}$ decay is slower than $t^{-1}$. This can be resolved by accounting for the interplay of the origin times in Eqs. (3). This means that $t^{-3/2}$ decay is only an intermediate regime, valid up to some large time $t_3$, when the increasing in time intervortex distance $\ell(t)=1/\sqrt{\mathcal{L}(t)} \propto t^{3/4}$ becomes of the order of the largest scale $\Delta$. We cannot observe this final stage of the decay with $\mathcal{L}(t) \sim 1\,\mathrm{cm}^{-2}$, due to a large noise level.

Nevertheless, and this is important for discussion in Sec. IV, the classical $t^{-3/2}$-fit (3b) describes the observed decays during tens of seconds after $t_2$, when $\mathcal{L}(t)$ decreases by more than two orders of magnitude. This is seen, e.g. in Fig. 2(a) for set # 3 with $T=1.35$ K and

$\mathcal{L}_0 \sim 10^6\,\mathrm{cm}^{-2}$.

### 2. Counterflow and pure superflow

Typical examples of decaying VLD in superfluid $^4$He after switching off the counterflow are shown in Figs. 3(a,b) and after switching off pure superflow – in Figs. 3(c,d). These data, obtained in 10 mm channel for $T = 1.45\,\mathrm{K}$, are very similar to the data from 7 mm channel (not discussed in this paper). The lines are marked as for the coflow, Figs. 2, according to initial VLD : $\mathcal{L}_0 \sim 10^4$ (blue lines), $\mathcal{L}_0 \sim 10^5$ (green lines) and $\mathcal{L}_0 \sim 10^6$ (red lines).

Similar to the coflow, the decay of $\mathcal{L}(t)$ the may be divided into three stages: (i) The initial quantum stage that agrees with the $t^{-1}$-fit (3a), in the time interval $0 \lesssim t \lesssim t_1$, during which $\mathcal{L}(t)$ monotonically falls from about $\mathcal{L}_0$ to $\mathcal{L}_1$. The fitting parameters $b_1$ and $\tau_1$ together with the ratios $\mathcal{L}_0/\mathcal{L}_1$ and time $t_1$ are given in Tab. I. (ii) The final classical stage that agrees with the monotonic decay, described by $t^{-3/2}$-fit (3b). It starts at time $t \simeq t_2$ from the VLD $\mathcal{L}_2$ and lasts for several tens of seconds. Table I presents parameters $b_2$ and $\tau_2$ together with the ratios $\mathcal{L}_1/\mathcal{L}_2$ and time $t_2$. (iii) The intermediate stage between the quantum and the classical one, that lasts from $t_1$ to $t_2$ and during which $\mathcal{L}(t)$ decays from $\mathcal{L}_1$ to $\mathcal{L}_2$. The most striking feature that qualitatively differs this stage in counterflow and pure superflow from that in coflow is the non-monotonic character of the decay, clearly seen in Figs. 3, especially for large $\mathcal{L}_0$.

Comparing Figs. 2 for coflow with Figs. 3 for counterflow and pure superflow, one sees that there are no qualitative differences between all these flows at the initial quantum and final classical stages of the monotonic decay of VLD. The same conclusion follows from Tab. I, which demonstrates only a small quantitative difference between these flows at these stages, perhaps with some scattered values of displayed parameters for all flows.

As we discussed above, the counterflow and pure superflow qualitatively differ from the coflow only at the intermediate stage, demonstrating non-monotonic decay. As we explain below, this difference is a consequence of a very different character of the steady-state energy spectra (energy distribution between scales) in the counterflow and pure superflow with non-zero values of the counterflow velocity $U_{\mathrm{ns}} \neq 0$ from the stationary spectrum in the coflow, for which $U_{\mathrm{ns}} = 0$. The energy spectra of quantum $^4$He turbulence for all three types of flows, discussed in our paper, are the subject of Sec. III.

## III. OVERVIEW OF THE ENERGY SPECTRA OF QUANTUM $^4$He TURBULENCE

In order to rationalize the discussed above time evolution of the VLD $\mathcal{L}(t)$ in counterflow, coflow and pure superflow turbulence we need to clarify the energy spectra in all these types of quantum turbulence. We begin in Sec. III A, with discussion of the energy spectrum of coflow grid turbulence. After that, following recent Ref. [23] we overview in Sec. III B the effect of the counterflow velocity on the steady-state energy spectra of the counterflow and pure superflow turbulence, important for the understanding of the appearance of a bump on the time dependence of $\mathcal{L}(t)$ at the intermediate stage of the decay.

### A. Steady-state coflow grid turbulence

In coflow turbulence, it is very natural to assume that the mean velocity profiles of the normal and superfluid velocity, $U_{\mathrm{n}}(y)$ and $U_{\mathrm{s}}(y)$, practically coincide almost everywhere in the channel, perhaps except for narrow regions near the walls. The reasons are the following: The pressure drop is the same for both fluid components, relatively large mutual friction tries to lock the mean normal and superfluid velocities and only kinematic viscosities, important in the near-wall region (viscous and buffer layers) are different: $\nu_{\mathrm{n}} \neq \nu_{\mathrm{s}} = 0$. This causes some decoupling of the velocities in that region. As a consequence, the kinetic energy is transferred from the mean flow to the turbulent velocity fluctuations. Its value is proportional to the shear of the mean velocity $S(y) = dU(y)/dy$ and is practically the same in the entire channel for the normal and superfluid components. Thus we assume that the energy spectra of the normal and superfluid components practically coincide in the entire energy containing and inertial intervals of scales,

$$\mathcal{E}_{\mathrm{n}}(k) = \mathcal{E}_{\mathrm{s}}(k)\,. \tag{6a}$$

Ignoring unessential for present discussion of intermittency effects, we can use classical K41 energy spectra for both $^4$He components:

$$\mathcal{E}_{\mathrm{n}}^{\mathrm{cl}}(k) = \mathcal{E}_{\mathrm{s}}^{\mathrm{cl}}(k) \simeq C_{\mathrm{K41}} \varepsilon^{2/3} k^{-5/3}\,, \tag{6b}$$

where $\varepsilon_{\mathrm{n}} = \varepsilon_{\mathrm{s}} \equiv \varepsilon$, are the corresponding energy fluxes over inertial interval of scales and $C_{\mathrm{K41}} \simeq 1.4$ is the Kolmogorov constant.

In the classical channel flow [28] the mean turbulent kinetic energy $E^{\mathrm{cl}}$ and the kinetic energy of the mean flow $E^{\mathrm{mf}} = \langle U \rangle_\Delta^2 /2$ are related by:

$$\frac{E^{\mathrm{cl}}}{E^{\mathrm{mf}}} \simeq \frac{4}{[\kappa_{\mathrm{K}}^{-1} \ln(\mathrm{Re}_\tau) + B_{\mathrm{cl}}]^2}\,. \tag{7}$$

Here Reynolds number $\mathrm{Re}_\tau = \Delta\sqrt{p'\Delta}/\nu_{\mathrm{n}}$, where $\Delta$ is half channel width, $p'$ is the pressure gradient that drives the flow, $\kappa_{\mathrm{K}} \approx 0.41$ is the von Kármán constant and the constant $B_{\mathrm{cl}} \approx 5$. Using this equation, we can find $E^{\mathrm{cl}}$, which is dominated by the outer scale of turbulence $\sim \Delta$. Then, using Kolmogorov energy spectrum (6b), we estimate its part, $E_\ell$, originating from the intervortex scale $\ell$: $E_\ell \simeq E^{\rm cl}(\ell/\Delta)^{2/3}$. This part corresponds to the energy of the random vortex tangle of VLD $\mathcal{L}$, that may be found from the relation $E_\ell \simeq \kappa^2 \mathcal{L}$. By equating these expressions we can roughly estimate VLD $\mathcal{L}_*$, connected to the classical energy spectrum for different values of $U_{\rm s}$. The result is the following: for smaller $U_{\rm s}$, that corresponds to $\mathcal{L} \sim 10^4\,{\rm cm}^{-2}$, the estimated value of VLD $\mathcal{L}_*$ is close or above the experimental value of initial VLD $\mathcal{L}_0$: $\mathcal{L}_* \simeq (0.9 \div 1.4)\, \mathcal{L}_0$, while for larger $U_{\rm s}$ (corresponding to $\mathcal{L} \sim 10^6\,{\rm cm}^{-2}$) we find that $\mathcal{L}_* \simeq (0.5 \div 1.0)\, \mathcal{L}_0$.

This means that, in order to explain how the observed initial value of VLD $\mathcal{L}_0$ may be larger than the classically generated value $\mathcal{L}_*$, one should find some additional mechanism of vortex generation, besides the classical flow instabilities in the channel flow.

It is almost common belief (see, e.g., Ref. [2,29]) that large scale motions of normal and superfluid components are dynamically locked, in the sense that their turbulent velocities coincide everywhere in space at any time:

$$\boldsymbol{v}_{\rm n}(\boldsymbol{r},t) = \boldsymbol{v}_{\rm s}(\boldsymbol{r},t)\,, \qquad (8)$$

and thus the counterflow mechanism of the vortex lines generation seems to be absent.

However, as was shown in detail in Refs. [30,31], the normal and superfluid energy spectra deviate from each other in the crossover region $k\ell \sim 1$. The relation (8) is also violated in the narrow region near the channel walls and near the surface of the grid at the channel entrance, where normal and superfluid components satisfy different boundary conditions. In these areas the velocities of components differ and excite a random vortex tangle leading to a peak in the energy spectrum near the crossover scale $k\ell \sim 1$.

We thus conclude that the resulting form of the turbulent energy spectrum $\mathcal{E}(k)$ may include, besides the classical Kolmogorov region (6b) (spanning scales from the integral length-scale $\Delta$ down to the intervortex distance $\ell$), also the quantum contribution $\mathcal{E}_{\rm s}^{\rm Q}(k)$, which has a form of a peak corresponding to the random tangle of quantized vortex lines. This peak has large-$k$ asymptote $\propto 1/k$ originating from the velocity field near the quantized vortex lines. In the region of $k < \pi/\ell$ the quantum peak is adjoined by the classical region with the thermodynamic equilibrium spectrum $\propto k^2$, describing equipartition of energy between degrees of freedom[32].

### B. Steady-state counterflow and pure superflow $^4$He turbulence

Detailed analysis of the steady-state energy spectra in the counterflow turbulence, that reflects our current understanding of this problem, will be done below in Sec. V. Here we overview only some earlier results on the counterflow energy spectra, which, nevertheless are sufficient to formulate in Sec. IV our model of decaying quantum turbulence.

#### 1. Counterflow induced decoupling in quantum turbulence

The first important ingredient of understanding of the steady-state energy spectrum in counterflow turbulence is decoupling of the normal and superfluid velocity fluctuations, $\boldsymbol{u}_{\rm n}(\boldsymbol{r},t)$ and $\boldsymbol{u}_{\rm s}(\boldsymbol{r},t)$, caused by the their sweeping in opposite direction by the corresponding mean velocities $\boldsymbol{U}_{\rm n}$ and $\boldsymbol{U}_{\rm s}$. Considering the velocity fluctuations of characteristic scale $R$ (referred for the shortness to as $R$-eddies) one can easily estimate their overlapping time as

$$\tau_{\rm ov}(R) = R/U_{\rm ns}\,, \quad \boldsymbol{U}_{\rm ns} = \boldsymbol{U}_{\rm n} - \boldsymbol{U}_{\rm s}\,. \qquad (9)$$

Another characteristic time in the problem of velocity coupling is the interaction time $\tau_{\rm int}$ of normal and superfluid eddies, due to the mutual-friction force acting between them. To estimate $\tau_{\rm int}$, we recall[33] that normal eddies act upon the superfluid eddies with the force (per unit mass) $\boldsymbol{F}_{\rm s}$

$$\boldsymbol{F}_{\rm s}(\boldsymbol{r},t) \simeq \Omega\left[\boldsymbol{U}_{\rm ns} + \boldsymbol{u}_{\rm n}(\boldsymbol{r},t) - \boldsymbol{u}_{\rm s}(\boldsymbol{r},t)\right]\,, \quad (10{\rm a})$$
$$\Omega = \alpha(T)\kappa\mathcal{L}\,,$$

where $\Omega$ is mutual friction frequency for the superfluid and $\alpha(T)$ is the dimensionless mutual friction parameter, tabulated in Ref. [27]. On the other hand, the superfluid eddies act on the normal fluid ones with the force

$$\boldsymbol{F}_{\rm n}(\boldsymbol{r},t) \simeq -\Omega_{\rm n}\left[\boldsymbol{U}_{\rm ns} + \boldsymbol{u}_{\rm n}(\boldsymbol{r},t) - \boldsymbol{u}_{\rm s}(\boldsymbol{r},t)\right]\,, \quad (10{\rm b})$$

where $\Omega_{\rm n} = \Omega\rho_{\rm s}/\rho_{\rm n}$. As required by the conservation of the mechanical momentum, the weighted sum $\rho_{\rm s}\boldsymbol{F}_{\rm s}(\boldsymbol{r},t) + \rho_{\rm n}\boldsymbol{F}_{\rm n}(\boldsymbol{r},t) = 0$, while the difference $\boldsymbol{F}_{\rm ns}(\boldsymbol{r},t) = \boldsymbol{F}_{\rm s}(\boldsymbol{r},t) - \boldsymbol{F}_{\rm n}(\boldsymbol{r},t)$ plays an important role in the intrinsic dynamics of the velocity differences. In the equation for $\boldsymbol{F}_{\rm ns}(\boldsymbol{r},t)$

$$\boldsymbol{F}_{\rm ns}(\boldsymbol{r},t) \simeq \Omega_{\rm ns}[\boldsymbol{U}_{\rm ns} + \boldsymbol{u}_{\rm n}(\boldsymbol{r},t) - \boldsymbol{u}_{\rm s}(\boldsymbol{r},t)]\,, \qquad (10{\rm c})$$
$$\Omega_{\rm ns} = \Omega\rho/\rho_n = \alpha_{\rm ns}\kappa\mathcal{L}\,, \quad \alpha_{\rm ns} = \alpha(T)\rho/\rho_{\rm n}(T)\,.(10{\rm d})$$

The resulting mutual friction parameter $\alpha_{\rm ns}$ is only weakly dependent on $T$, varying between 0.7 and 0.5 in the temperature range $(1.4 \div 1.9)\,{\rm K}$ [27]. The first term in Eq. (10c), $\Omega_{\rm ns}\boldsymbol{U}_{\rm ns}$, is compensated in the stationary counterflow by the force, originating from the temperature drop. The second term, $\Omega_{\rm ns}\left[\boldsymbol{u}_{\rm n}(\boldsymbol{r},t) - \boldsymbol{u}_{\rm s}(\boldsymbol{r},t)\right]$, responsible for the coupling of the velocities by mutual friction, is proportional to the scale-independent "interaction frequency" $\Omega_{\rm ns}$. Accordingly, the mutual-friction interaction time can be estimated as follows:

$$\tau_{\rm int} = \frac{\pi}{\Omega_{\rm ns}} = \frac{\pi}{\alpha_{\rm ns}\kappa\mathcal{L}}\,. \qquad (11)$$

As was shown in Ref. [23] (see also[29]), the overlapping time $\tau_{\rm ov}$ (9) should be compared with the $R$-independent interaction time (11). If, for large $U_{\rm ns}$ or small $R$, the overlapping time is much smaller than the interaction time, $\tau_{\rm ov} \ll \tau_{\rm int}$, the mutual friction does not have enough time to couple the normal and superfluid velocity fluctuations. In this case one expects that the velocities remain uncoupled. Otherwise, i.e., for large scales $R$ or small $U_{ns}$, $\tau_{ov} \gg \tau_{int}$, and one should expect full coupling (8).

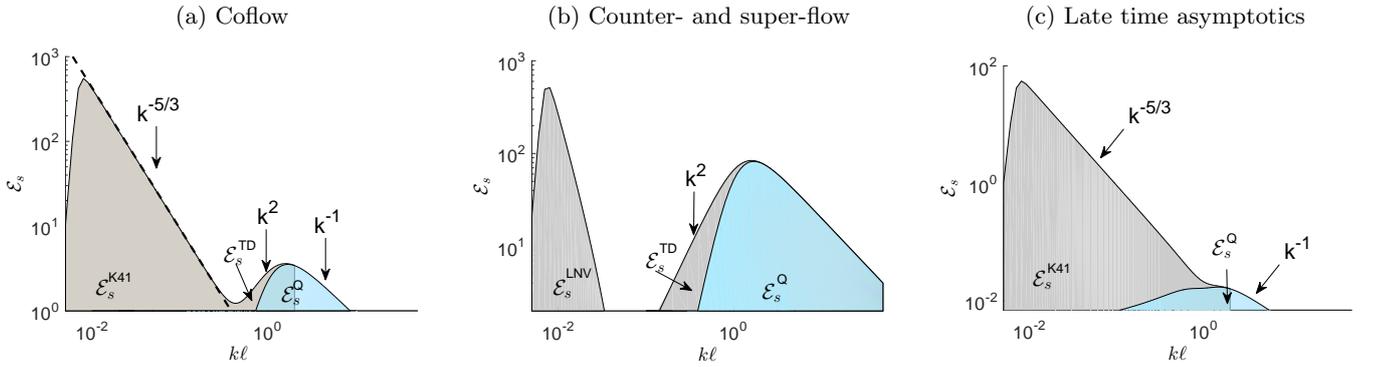

FIG. 4: (color online) Sketch of the stationary superfluid turbulent energy spectrum in log-log coordinates, $\log \mathcal{E}_s(k)$ vs. $\log(k\ell)$. According to Eq. (4) spectrum $\mathcal{E}_s(k)$ consists of classical $\mathcal{E}_s^{cl}(k)$ and quantum $\mathcal{E}_s^Q(k)$ parts, colored in light blue and cyan. In coflow (Panel a) $\mathcal{E}_s^{cl}(k)$ consists of cascade part $\mathcal{E}_s^{K41}(k) \propto k^{-5/3}$ (for $k < k_\times$) and thermodynamic equilibrium part $\mathcal{E}_s^{TD}(k) \propto k^2$ for $1/\ell \gtrsim k > k_\times$. In counterflow and pure superflow (Panel b) the quantum contribution $\mathcal{E}_s^Q(k)$ and the classical thermal bath part $\mathcal{E}_s^{TD}(k)$ look similarly to that in coflow, while the cascade part, supercritical LNV-spectrum, $\mathcal{E}_{LNV}$ given by Eqs. (19), ends at some $k_* < 1/\ell$ and does not provide energy to the quantum vortex tangle in stationary regime. After switching off the counterflow or pure superflow, the spectrum shown on Panel b evolves to that shown in Panel c, switching on the energy flux toward quantum vortex tangle after some delay.

Therefore the coupling-decoupling process is governed by the dimensionless "decoupling" parameter

$$\zeta(k) = \frac{\tau_{int}}{\tau_{ov}} \simeq \frac{kU_{ns}}{\Omega_{ns}}, \quad \text{with } k \approx \pi/R. \quad (12a)$$

The analytical theory of the coupling-decoupling processes[23], using Langevin inspired approach to model the nonlinear term, results in the analytical expression for the cross-correlation $\mathcal{E}_{ns}(k)$ in terms of $\mathcal{E}_s(k)$ and $\mathcal{E}_n(k)$, defined by Eqs. (B5). Using Eqs. (13) and (16) from Ref. [23], the result in the relevant for our experiments range of parameters can be written as:

$$\mathcal{E}_{ns}(k) = D(k)\frac{\Omega \mathcal{E}_n(k) + \Omega_n \mathcal{E}_s(k)}{\Omega_{ns}}, \quad \Omega_{ns} = \Omega + \Omega_n, (12b)$$

$$D(k) = \frac{1}{\zeta(k)} \arctan[\zeta(k)]. \quad (12c)$$

In Eqs. (12) the dimensionless "decoupling function" $D(k)$ is defined via the dimensionless "decoupling parameter" $\zeta(k)$, and describes the decoupling of the normal and superfluid velocity fluctuations, caused by the counterflow velocity. The plot of $D$ as a function of $k/k_\times$ is shown in Fig. 5, where

$$k_\times \simeq 2\,\Omega_{ns}/U_{ns}, \quad (12d)$$

The analysis of Eqs. (12), given in Ref. [23], allows us to conclude that the normal and superfluid velocity fluctuations are coupled in the wave number interval from the smallest one, $k_0 \simeq \pi/\Delta$, up to $k_\times$, which is slightly larger than $k_0$ and are decoupled in the rest of $k$-range up to $\pi/\ell \gg k_\times$.

2. *Qualitative analysis of the counterflow energy spectrum*

The next question of principal importance for the analysis of our experiments is how the decoupling of the velocities at small scales affects the steady-state energy spectra before switching off the flows. We will discuss this problem in details later in Sec. V. Nevertheless, to get a qualitative form of the energy spectra, sketched in Fig. 4, required for formulation of our model of de-caying quantum turbulence, it is sufficient to overview relatively resent results on energy spectra in turbulent $^3$He, see Refs. [33,34] and analysis, presented in Ref. [23].

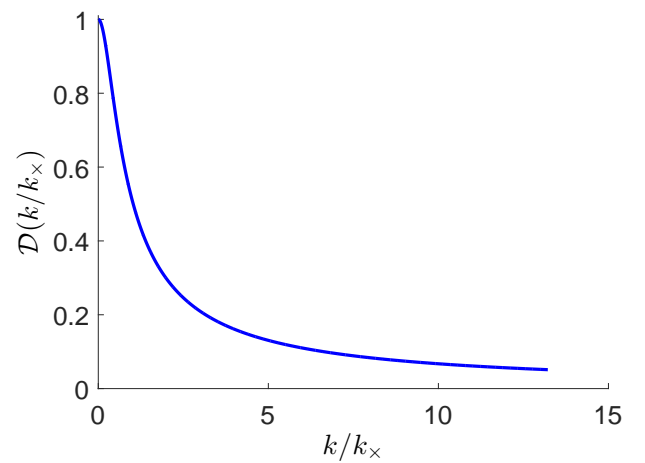

FIG. 5: (color online) Decoupling function $D(k/k_\times) = \mathcal{E}_{ns}/\mathcal{E}_{ns}^{(0)}$ vs dimensionless wave number $k/k_\times$.

The coupling – decoupling process of normal and superfluid velocity fields crucially affects the energy dissipation due to mutual friction and, correspondingly, modifies the energy spectra $\mathcal{E}_{\rm s}(k,t)$ and $\mathcal{E}_{\rm n}(k,t)$. To demonstrate this, we use the evolution equations for these objects, derived in Ref. [23]:

$$\frac{\partial \mathcal{E}_{\rm s}(k,t)}{2\partial t} + \mathcal{NL}_{\rm s} = \Omega\big[\mathcal{E}_{\rm ns}(k,t) - \mathcal{E}_{\rm s}(k,t)\big], \quad (13a)$$

$$\frac{\partial \mathcal{E}_{\rm n}(k,t)}{2\partial t} + \mathcal{NL}_{\rm n} = \Omega_{\rm n}\big[\mathcal{E}_{\rm ns}(k,t) - \mathcal{E}_{\rm n}(k,t)\big]. (13b)$$

Here $\mathcal{NL}_{\rm s,n}$ are nonlinear terms, which conserve kinetic energy and therefore may be presented in the divergent form:

$$\mathcal{NL}_{\rm s} = \frac{d\varepsilon_{\rm s}}{dk}, \quad \mathcal{NL}_{\rm s} = \frac{d\varepsilon_{\rm s}}{dk}. \quad (13c)$$

Here $\varepsilon_{\rm s}(k)$ and $\varepsilon_{\rm n}(k)$ are the energy fluxes in corresponding subsystems. For $k \gg k_\times$, due to the decoupling, $\mathcal{E}_{\rm ns}(k) \ll \mathcal{E}_{\rm s}(k)$. Therefore the term $\mathcal{E}_{\rm ns}(k)$ may be neglected in the RHS of Eq. (13a), which becomes $-\Omega_{\rm s}\mathcal{E}_{\rm s}$. This is similar to the equation for $\mathcal{E}_{\rm s}$ for the quantum turbulence in superfluid $^3$He-B, written below for the steady-state case:

$$\frac{d\varepsilon_{\rm s}(k)}{dk} + \Omega\,\mathcal{E}_{\rm s} = 0\,. \quad (14a)$$

Notice that the energy flux over scale $\varepsilon(k)$ can be expressed in the terms of the third-order correlation function of $\boldsymbol{u}_{\rm s}(k)$ exactly as in classical turbulence. In order to proceed further, one can borrow a closure procedure from classical turbulence that expresses $\varepsilon_{\rm s}(k)$ in terms of the energy spectrum $\mathcal{E}_{\rm s}(k)$. Even though this step is widely used, it is worth remembering that it is uncontrolled. The simplest algebraic closure relation suggested by Kovasznay [35]

$$\varepsilon_{\rm s}(k) \simeq \frac{5}{8} k^{5/2} \mathcal{E}_{\rm s}^{3/2}(k)\,, \quad (14b)$$

just follows from the K41 dimensional reasoning

$$\mathcal{E}_{\rm s}(\varepsilon|k) = C_{\rm K}\varepsilon^{2/3} k^{-5/3}, \quad C_{\rm K} = (8/3)^{2/3} \approx 1.4\,. \quad (14c)$$

The prefactor $\frac{5}{8}$ is chosen to simplify the appearance of some of the equations below and to get numerical value of the Kolmogorov constant $C_{\rm K}$ reasonably close to its experimental value.

Ordinary differential Eqs. (14)

$$\frac{5}{8}\frac{d}{dk}\big[k^{5/2}\mathcal{E}_{\rm s}^{3/2}(k)\big] + \Omega\,\mathcal{E}_{\rm s}(k) = 0\,, \quad (15)$$

should be solved with the boundary conditions $\mathcal{E}_{\rm s}(k_0) = \mathcal{E}_0$ at the lowest wave number $k_0$ in the inertial interval. The physical solution, found by L'vov, Nazarenko, Volovik (LNV) [33] in current notations is:

$$\mathcal{E}(k) = \mathcal{E}_0 \Big(\frac{k_0}{k}\Big)^3 \bigg[\Big(1 - \frac{\Omega}{\Omega_{\rm cr}}\Big)\Big(\frac{k}{k_0}\Big)^{2/3} + \frac{\Omega}{\Omega_{\rm cr}}\bigg]^2, (16a)$$

$$\Omega_{\rm cr} = \frac{5}{4}\sqrt{k_0^3\,\mathcal{E}_0}\,. \quad (16b)$$

At $\Omega = \Omega_{\rm cr}$ it becomes the scale-invariant "critical" spectrum

$$\mathcal{E}_{\rm cr}(k) = \mathcal{E}_0 \Big(\frac{k_0}{k}\Big)^3, \quad \text{LNV-critical.} \quad (17)$$

For $\Omega < \Omega_{\rm cr}$ solution (16a) can be considered as "subcritical" and written as follows:

$$\mathcal{E}_{\rm sb}(k) = \mathcal{E}_0 \frac{k_0^3}{k^{5/3}}\Big[\frac{1}{k^{2/3}} + \frac{1}{k_{\rm cr}^{2/3}}\Big]^2, \text{ LNV-subcritical,}$$

$$k_{\rm cr} = k_0\Big(\frac{\Omega}{\Omega_{\rm cr} - \Omega}\Big)^{3/2}. \quad (18a)$$

If $\Omega$ is close to $\Omega_{\rm cr}$ from below, $\Omega_{\rm cr} - \Omega \ll \Omega_{\rm cr}$, and $k$ is smaller than the crossover wave number $k_{\rm cr}$, $\mathcal{E}_{\rm sb}(k)$ is close to the critical solution (17). However at $k \gg k_{\rm cr}$, the subcritical spectrum (16a) has K41 asymptote

$$\mathcal{E}_{\rm sb}(k) \Rightarrow \mathcal{E}_0\Big(\frac{\varepsilon_\infty}{\varepsilon_0}\Big)^{2/3}\Big(\frac{k_0}{k}\Big)^{5/3},\ \varepsilon_\infty = \varepsilon_0\Big(1 - \frac{\Omega}{\Omega_{\rm cr}}\Big)^3, \quad (18b)$$

with the energy flux $\varepsilon_\infty$ smaller than the energy influx $\varepsilon_0$. The difference $\varepsilon_0 - \varepsilon_\infty$ is dissipated by the mutual friction.

In the case $\Omega > \Omega_{\rm cr}$, Eq. (16a) can be considered as "supercritical" spectrum

$$\mathcal{E}_{\rm sp}(k) = \mathcal{E}_0 \frac{k_0^3}{k^{5/3}}\Big[\frac{1}{k^{2/3}} - \frac{1}{k_*^{2/3}}\Big]^2, \text{ LNV-supercritical,}$$

$$k_* = k_0\Big(\frac{\Omega}{\Omega - \Omega_{\rm cr}}\Big)^{3/2}. \quad (19)$$

that terminates at some $k = k_*$. However, if $\Omega$ is close to $\Omega_{\rm cr}$ from above, $\Omega - \Omega_{\rm cr} \ll \Omega_{\rm cr}$, such that $k_* \gg k_0$, and $k < k_*$, we see that the supercritical spectrum is close to the critical one. Examples of three versions of the LNV spectra are plotted in Fig. 8. Note in passing that similar conclusions on behavior of $^3$He-B quantum turbulence follow from independent analysis of Vinen[36].

All this means that, assuming full decoupling of the velocity fields, the behavior of the superfluid component of $^4$He in the counterflow becomes similar to that in $^3$He-B turbulence with the normal fluid component at rest. Therefore one can expect that the energy distribution between scales for $k \gg k_{\rm cr}$ may be described by the LNV-spectrum (16) or similar localized spectrum.

For $k < k_\times$, due to partial velocity correlations, the energy dissipation due to mutual friction is much weaker than for $k > k_\times$, although it cannot be completely neglected as in coflowing $^4$He (with classical K41 energy spectrum). Thus we can expect only moderate suppression of the energy spectrum as compared to the K41 case, as was recently observed in Ref.[37] and analysed below in Sec. V. The resulting energy spectrum is sketched in Fig. 4(b). Besides the classical part, this spectrum includes a quantum peak, presumably more intensive than the peak in the coflow, shown in Fig. 4(a). These peaks have similar structure: large $k$-asymptotics $\propto 1/k$ and small $k$-asymptotics $\propto k^2$.

## IV. BASIC AND IMPROVED MODELS OF DECAYING QUANTUM TURBULENCE

In this Section, we first propose and solve a basic model of the VLD evolution, $\mathcal{L}(t)$, in decaying quantum turbulence. The model will then be further developed to clarify the experimental facts in more detail.

### A. Basic model of decaying coflow turbulence

Assume that the time derivative of $\mathcal{L}(t)$ consists of a simple sum of the quantum decay term $-\eta_{\rm qn}$ and the classical source term $\eta_{\rm cl}$, neglecting possible processes of their interaction:

$$\frac{d\mathcal{L}(t)}{dt} = -\eta^{\rm Q} + \eta^{\rm cl} . \qquad (20)$$

The quantum tangle decay (without counterflow velocity) is usually discussed in the framework of the Vinen equation[14,20] with the (quantum) decay term $\eta^{\rm Q} = \chi_2 \kappa \mathcal{L}^2/(2\pi)$. Estimating the phenomenological coefficient $\chi_2$ in the local induction approximation[38,39], we rewrite $\eta^{\rm Q}$ in terms of the dimensionless mutual friction parameter $\alpha$ and, the vortex line curvature $\widetilde{S}$, normalized by the intervortex distance $\ell$, $c_2 \sim 1$:

$$\eta^{\rm Q} \simeq \alpha\,\kappa\,c_2^2 \frac{\Lambda}{4\pi} \mathcal{L}^2 , \quad c_2 \equiv \widetilde{S}\ell, \quad \Lambda = \ln(\ell/a_0) . \quad (21{\rm a})$$

Bearing in mind that for our conditions the parameter $\Lambda/(4\pi) \simeq 1$ varies from 0.9 to 1.1 and $c_2 \simeq 1$ depends weakly on temperature[41], we can simplify this relation by introducing a dimensionless fitting parameter $d_1$, and write

$$\eta^{\rm Q} = \frac{\alpha\,\kappa}{d_1} \mathcal{L}^2 . \qquad (21{\rm b})$$

With this $\eta^{\rm Q}$ and without classical source term $\eta^{\rm cl}$, Eq. (20) has well known $t^{-1}$-solution (3a) with

$$b_1 = d_1/(\alpha\,\kappa \mathcal{L}_0 \tau_1) . \qquad (22)$$

The positive contribution $\eta^{\rm cl}$ in the RHS of Eq. (20) originates from the direct energy flux $\varepsilon$ from the classical energy containing scale $k\Delta \sim 1$, to the quantum energy peak $\mathcal{E}^{\rm Q}(k)$, located at $k\ell \sim 1$, see Fig. 4. To estimate $\eta^{\rm cl}$, recall that $\mathcal{E}^{\rm Q} \sim \kappa^2 \mathcal{L}$ (see also Ref.[40]). Therefore the flux of VLD from the classical scales to the quantum energy peak may be estimated as

$$\eta^{\rm cl} \simeq \varepsilon/\kappa^2 . \qquad (23)$$

Integrating Eq. (6b), we find the total classical kinetic energy

$$\mathcal{E}^{\rm cl} = \int \mathcal{E}^{\rm cl}(k)\,dk \simeq C_{\rm K41} \varepsilon^{2/3} \int_{\pi/\Delta}^{\pi/\ell} k^{-5/3} dk$$

$$= \frac{3\,C_{\rm K41}\varepsilon^{2/3}}{2\pi^{2/3}} \big[\Delta^{2/3} - \ell^{2/3}\big] \simeq \Big(\frac{\varepsilon\,d_2 \Delta}{2}\Big)^{2/3} . \quad (24{\rm a})$$

Here we roughly estimated lower and upper limits of the inertial interval as $\pi/\Delta$ and $\pi/\ell$. Bearing in mind that at our experimental conditions $\Delta$ significantly exceeds $\ell$, we neglected $\ell$ with respect to $\Delta$. Finally, we replaced in Eq. (24a) all numerical factors by a new fitting dimensionless constant $d_2 \simeq 1$. As a result

$$\varepsilon \simeq 2\,(\mathcal{E}^{\rm cl})^{3/2}/(d_2 \Delta) . \qquad (24{\rm b})$$

Now we can write the classical energy balance equation

$$\frac{d\mathcal{E}^{\rm cl}}{dt} = -\varepsilon = -2 \frac{(\mathcal{E}^{\rm cl})^{3/2}}{d_2 \Delta} , \qquad (24{\rm c})$$

with the solution:

$$\mathcal{E}^{\rm cl}(t) = \Big(\frac{d_2 \Delta}{t - \tau_2}\Big)^2 , \quad \tau_2 = \frac{d_2 \Delta}{\sqrt{\mathcal{E}_0^{\rm cl}}} , \qquad (24{\rm d})$$

where $\mathcal{E}_0^{\rm cl}$ is the initial energy at $t = 0$. Together with Eqs. (24b) and (23) this finally gives

$$\eta^{\rm cl}(t) = \frac{\varepsilon^{\rm cl}(t)}{\kappa^2} = \frac{2\big(d_2 \Delta\big)^2}{\big(t - \tau_2\big)^3} . \qquad (24{\rm e})$$

Collecting Eqs. (20), (21b), (24d) and (24e), we finally suggest a simple model of the VLD decay in the form:

$$\frac{d\mathcal{L}}{dt} = \frac{2\,(d_2 \Delta)^2}{\kappa^2 (t - \tau_2)^3} - \frac{\alpha\,\kappa}{d_1} \mathcal{L}^2 . \qquad (25)$$

Here $d_1$ and $d_2$ are dimensionless phenomenological parameters. Note that $\tau_2$ is of the order of turnover time $\tau_\Delta \simeq \Delta/\sqrt{\mathcal{E}_0^{\rm cl}}$ of the largest (energy containing) eddies in the flow.

When the classical energy $\mathcal{E}_0^{\rm cl}$ is small, the virtual origin time $\tau_2$, according to Eq. (24d), is large and the first term in the RHS of Eq. (25) may be neglected. In this case Eq. (25) may be easily solved, giving $t^{-1}$-decay $\mathcal{L}^{\rm Q}$, Eq. (3a). Assume now that for large times the time derivative $d\mathcal{L}/dt$ in Eq. (25) may be neglected. Then Eq. (25) reproduces $t^{-3/2}$-decay $\mathcal{L}^{\rm cl}$, Eq. (3b) with

$$d_2 = b_2 \frac{(\kappa\,\tau_2)^{3/2}}{\Delta} \mathcal{L}_0 \sqrt{\frac{\alpha}{2 d_1}} , \qquad (26{\rm a})$$

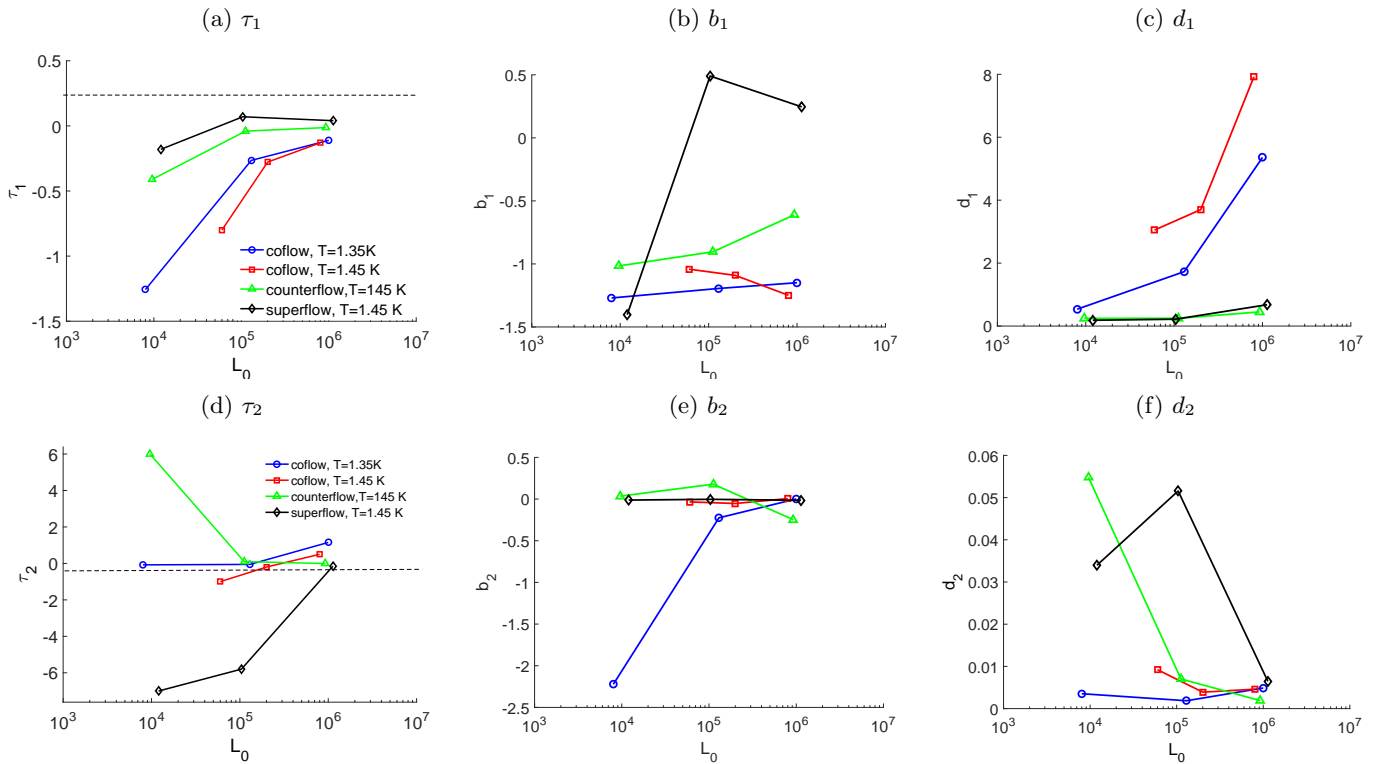

FIG. 6: (color online) Virtual origin times $\tau_1$ and $\tau_2$ for the quantum (3a) [panel (a)] and classical (3b) decay [panel (d)] vs initial VLD $\mathcal{L}_0$ in initial coflow, counterflow and pure superflow. Panels (b) and (e) display $\mathcal{L}_0$-dependence of the $b_1$ and $b_2$ fit parameters. Panels (c) and (f) display $\mathcal{L}_0$-dependence of the $d_1$ and $d_2$ fit parameters. The numerical values of all these parameters are given in Tab. I.

or by using $d_1$ from Eq. (22)

$$d_2 = b_2 \frac{\kappa \tau_2^{3/2}}{\Delta} \sqrt{\frac{\mathcal{L}_0}{2 b_1 \tau_1}} \ . \quad (26b)$$

Neglecting for simplicity the time dependence of the first term in the RHS of Eq. (25), one can solve it exactly with the result:

$$\mathcal{L}(t) = \mathcal{L}^{\mathrm{cl}}(t) \coth\left[\frac{\mathcal{L}^{\mathrm{cl}}(t)}{\mathcal{L}^{\mathrm{Q}}(t)}\right] \ . \quad (27)$$

Remind that

$$\lim_{x \to 0} \coth[x] \to 1/x, \ \lim_{x \to \infty} \coth[x] \to 1 \ .$$

Therefore, when $\mathcal{L}^{\mathrm{Q}} \gg \mathcal{L}^{\mathrm{cl}}$, $\mathcal{L}(t) \to \mathcal{L}^{\mathrm{Q}}$. Otherwise, in the limit $\mathcal{L}^{\mathrm{Q}} \ll \mathcal{L}^{\mathrm{cl}}$, $\mathcal{L}(t) \to \mathcal{L}^{\mathrm{cl}}$. For $\mathcal{L}^{\mathrm{Q}} \sim \mathcal{L}^{\mathrm{cl}}$, the function $\mathcal{L}(t)$ describes a smooth transition between $\mathcal{L}^{\mathrm{Q}}$ and $\mathcal{L}^{\mathrm{cl}}$, being always larger than both $\mathcal{L}^{\mathrm{Q}}$ and $\mathcal{L}^{\mathrm{cl}}$. In this way, the approximate solution (27) for $\mathcal{L}(t)$, interpolates the solution of the model Eq. (25) between the small- and large-time asymptotes.

Even better interpolation between exact asymptotes $\mathcal{L}^{\mathrm{Q}}$ and $\mathcal{L}^{\mathrm{cl}}$ gives the following modification of Eq. (27):

$$\bar{\mathcal{L}}(t) = \mathcal{L}^{\mathrm{cl}}(t) \coth\left[\frac{\mathcal{L}^{\mathrm{cl}}(0)}{\mathcal{L}^{\mathrm{Q}}}\right] \ . \quad (28)$$

Bearing in mind the approximate character of the evolution model Eq. (25), we will consider Eq. (28) for $\bar{\mathcal{L}}(t)$ as an analytical form of the VLD time dependence $\mathcal{L}(t)$, practically equivalent to suggested basic model of decaying superfluid turbulence.

### B. Analysis of the decay fitting parameters

The $\mathcal{L}_0$-dependences of the fitting parameters $\tau_1$ and $\tau_2$, $b_1$ and $b_2$, $d_1$ and $d_2$ for three different types of the flow are given in Tab. I. For clarity, we additionally display these dependencies in Figs. 6.

#### 1. Quantum and classical origin times $\tau_1$ and $\tau_2$

The dependence of the quantum origin times $\tau_1$ on the initial VLD $\mathcal{L}_0$ for coflow, counterflow and pure superflow is shown in Fig. 6(a). As expected from Eqs. (3a) and (22), the values of $\tau_1$ for coflow are negative and monotonically increase with $\mathcal{L}_0$. There values of $\tau_1$ for coflow are clearly different from those in the counterflow and pure superflow, especially for smaller $\mathcal{L}_0$. This is related to the fact that in the coflow, the classical energy flux towards the quantum vortex tangle is present from the very beginning of the decay (at $t=0$), while in the counterflow and pure superflow it appears only after some delay, required for developing the Kolmogorov cascade from the initial spectrum sketched in Fig. 4(b).

As expected, the classical origin time $\tau_2$ tends to decrease for larger $\mathcal{L}_0$, as depicted in Fig. 6(d). The most striking fact, seen in Fig. 6(d), is that $\tau_2$ can be positive. This definitely contradicts the basic model of $\mathcal{L}(t)$-decay, formulated in Sec. IV A. Non-monotonic evolution of $\mathcal{L}(t)$, clearly seen in Figs. 3, also calls for improvements of the basic model of $\mathcal{L}(t)$-decay, in order to account for the time delay of the energy flux into the quantum vortex tangle. This is done in Sec. IV D, prefaced by Sec. IV C, devoted to the study of the energy-flux delay in the decaying turbulence in counterflow and pure superflow, caused by the later development of the Richardson-Kolmogorov cascade from the localized energy spectra.

### 2. Quantum and classical parameters $b_1$, $d_1$ and $b_2$, $d_2$

The $\mathcal{L}_0$-dependencies of the fit parameters $b_1$, $d_1$ and $b_2$, $d_2$ are shown in Figs. 6(b, c) and Figs. 6(e, f).

Quantum parameters $b_1$, $d_1$ are of the order of unity; their deviation from unity reflects the non-universal character of the transient regime after switching off the flow: the mean velocity has to relax to zero, anisotropic statistics of the vortex tangle, affected by the mean flow, has to become isotropic, etc. We will not discuss these complicated issues in this paper, as they do not seem to contribute to better understanding of the basic physical picture of decaying quantum turbulence in three types of the flow.

The same can be said about the scatter of values of the classical fitting parameters $b_2$, $d_2$ for different types of the flow. Notice only that their values, being much smaller than unity, become even smaller for larger $\mathcal{L}_0$. The same tendency is demonstrated by the ratio $\mathcal{L}_2$ (from which the classical decay begins) to the initial value of VLD $\mathcal{L}_0$. This may be easily interpreted in the following way: the random vortex tangle decays according to the quantum $t^{-1}$ law as long as the energy influx from the classical part of superfluid turbulence may be neglected. This holds as long as the tail of classical energy spectrum $\mathcal{E}_{\rm s}^{\rm K41}(\pi/\ell)$ is smaller than the quantum energy spectrum $\mathcal{E}_{\rm s}^{\rm Q}(\pi/\ell)$ at this scale. Therefore the ratio $\mathcal{E}_{\rm s}^{\rm Q}(\pi/\ell)/\mathcal{E}_{\rm s}^{\rm K41}(\pi/\ell)$ may be roughly estimated as $1/b_2 \gg 1$ or as a value between the ratios $\mathcal{L}_0/\mathcal{L}_1$ and $\mathcal{L}_0/\mathcal{L}_2$ (remind that $\mathcal{L}_1$ is the VLD at which the quantum decay terminates). As shown in Tab. I, both ratios are much larger than unity and both tend to increase for larger $\mathcal{L}_0$.

Based on this analysis, we conclude that the quantum peak of energy at the intervortex scale $\ell$, as a rule, dominates over the tail of classical energy spectrum at this scale, as depicted in Figs. 4. Notice that, generally speaking, this qualitative conclusion may be guessed just from the observation of the decay laws, shown in Figs. 2. What is added by our analysis is a semi-quantitative estimate of the ratio $\mathcal{E}_{\rm s}^{\rm Q}(\pi/\ell)/\mathcal{E}_{\rm s}^{\rm K41}(\pi/\ell)$.

### C. Energy-flux delay in the decaying counterflow turbulence

It is commonly accepted that the Richardson-Kolmogorov cascade develops from any localized in the $k$-space initial state over finite time, of the order of a few turnover times of the energy-containing eddies. Nevertheless, the details of the transient regime and how they depend on the initial state are poorly understood. In order to clarify the law of delay in "switching on" the energy flux $\varepsilon_{\rm cl}(t)$, Eq. (24e), that contributes to the RHS of the basic model (25), we adopt in our paper so-called Sabra-shell model of turbulence, successfully utilized in studies of quantum turbulence, e.g., in Refs. [31,34,42].

The required version of the Sabra model and the numerical procedure are described in Appendix A. Here we present only results of the time evolution of the energy spectrum $E_{\rm s}(k_m,t) = \langle |u_m|^2 \rangle / k_m$ in the decaying quantum turbulence from five types of the initial conditions $E_{\rm s}(k,0) \equiv E_0(k)$, shown in Fig. 7:

ICa – K41 energy spectrum (6b), $E_{0,a}(k) \propto k^{-5/3}$;

ICb – Experimental counterflow spectrum, $E_{0,b}(k) \propto k^{-2}$, reported in Ref. [37];

ICc – Critical LNV-spectrum (17) $E_{0,c}(k) \propto k^{-3}$;

ICd – Supercritical LNV-spectrum (19) $E_{0,d}(k) = 0$ for $k > k_*$;

ICe – Subcritical LNV-spectrum (18) $E_{0,e}(k)$.

The time dependence of the total energy for five types of the initial conditions, (a)-(e), are shown in Fig. 8(a). As expected, in all cases the large time asymptotics, $E(t) \propto t^{-2}$ agrees with Eq. (24d) – see black dashed line. What is important for current discussion, is the small time behavior for $t$ below few $\tau_\Delta$, shown in the inset of Fig. 8(a). Clearly, for critical and supercritical initial conditions, ICc and ICd, the energy $\mathcal{E}(t)$ (shown by coinciding solid red and dashed light blue lines) does not decay up to $t \approx \tau_\Delta$. This is the time required for development of the Richardson-Kolmogorov cascade, transferring energy into the dissipative range of scales (large $k$).

On the contrary, for the K41 initial condition, ICa, the energy decays from the very beginning – see blue solid line in the inset of Fig. 8(a). This behavior is also expected. For two intermediate initial conditions, ICb and ICe, the initial energy is also localized in the region of small $k$, but not so strongly, as in ICc and ICd cases. Accordingly, the time dependence of $E(t)$ for ICb and ICe cases, shown by green and orange lines, demonstrate intermediate behavior.

All these features are clearly seen in Fig. 8(b), showing the time dependence of the rate of energy dissipation

$\varepsilon(t)$. In our simulation we used a very small value of the kinematic viscosity, therefore $\varepsilon(t)$ is actually a measure of the energy flux via crossover wave number $\sim 1/\ell$ that serves as the energy flux from classical to quantum scale range. Thus, according to Eq. (23), we can say that the time dependence $\varepsilon(t)$ actually gives the time dependence of the important ingredient of the model Eq. (20), the classical energy source $\eta^{\rm cl}(t) \simeq \varepsilon(t)/\kappa^2$.

All plots of $\eta^{\rm cl}(t)$ in Fig. 8(b) have $t^{-3}$ asymptote for large $t$, in agreement with Eq. (24e). Moreover, for the K41 initial conditions, ICa, expected in coflow decay, and shown by the blue line, Eq. (24e) provides reasonable fit of $\eta^{\rm cl}(t)$ for all times, as shown by the solid black line in Fig. 8. Thus our numerical simulations support the basic model (25) of the decay in coflow, formulated in Sec. IV A.

The situation is completely different for other initial conditions, expected for the counterflow and pure superflow cases. There is the most striking difference for the well localized, critical and supercritical ICc and ICd, shown in Fig. 8 by (practically coinciding) red and light blue lines. One sees that for small time $\eta^{\rm cl}(t) \approx 0$, then it is sharply switching on and after few turnover times $\tau_\Delta$ reaches the "basic" behavior (24e). This can be accounted for by "improving" the basic model, introducing into Eq. (24e) for the classical source term the time delay function $F_{\rm del}(t)$:

$$\eta^{\rm cl} = \frac{2(d_2\,\Delta)^2}{(t+\tau_{\rm del}-\tau_2)^3} \Rightarrow F_{\rm del}(t)\,\frac{2(d_2\,\Delta)^2}{(t-\tau_2)^3}\ . \quad (29a)$$

It is convenient to choose $F_{\rm del}(t)$ as a square of new function $f_n(t)$ that has a simple form:

$$F_{\rm del}(t) = f_n^2(t)\,, \quad f_n(t) = \frac{t^n}{t^n + \tau_{\rm del}^n}\ . \quad (29b)$$

As required, $f_n(t) \to 1$ for $t \to \infty$. Generally speaking, $f(0) = 0$ only for the supercritical case, ICd. For all other cases (except for the K41 spectrum, which we are not discussing here) $f(0) \neq 0$, although small. This difference is not important for us and for simplicity we adopted in Eq. (29b) a simple assumption that $f_n(0) = 0$. The delay time $\tau_{\rm del}$ in Eq. (29b) is expected to be about the largest eddies turnover time, $\tau_\Delta$. Indeed, as seen in Fig. 11, $\tau_{\rm del} \simeq 0.4\tau_\Delta$.

Notice that the parameter $n$ in Eq. (29b), responsible for the sharpness of the delay function, is different for different initial conditions. Figure 11 demonstrates that for the weakly localized initial conditions, ICb and ICe, a reasonable approximation to the numerical observation can be reached with $n = 1$, while for strongly localized initial conditions a good fit corresponds to $n = 6$.

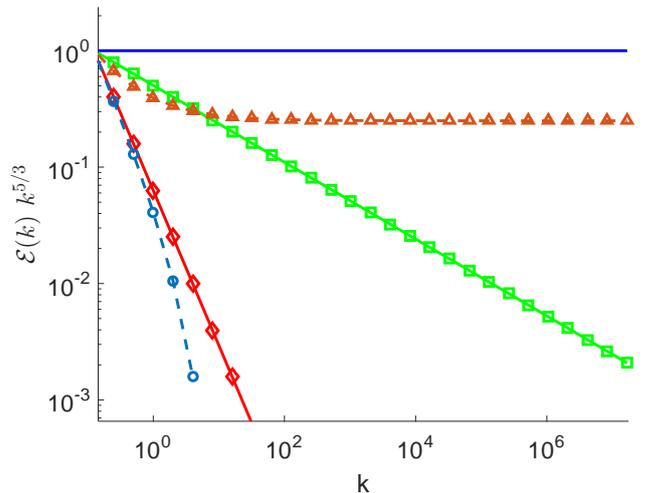

FIG. 7: (color online) Five variants of the stationary energy spectra, compensated by $k^{5/3}$, $k^{5/3}E_s(k)$, serving as the initial conditions for the Sabra-decay Eq. (A2): a) K41 energy $E_{0,a} \propto k^{-5/3}$, Eq. (6b) (solid blue line); b) Experimental counterflow spectrum, $E_{0,b}(k) \propto k^{-2}$, reported in[37]; (green line with squares) c) Critical LNV-spectrum (17) $E_{0,d}(k) \propto k^{-3}$ (red line with diamonds); d) supercritical LNV spectrum $E_{0,d}(k) = 0$ for $k > k_*$, Eq. (19); (light blue line with circles) e) subcritical LNV spectrum (18) (brown line with triangles).

### D. Improved model of VLD decay vs experiment

#### 1. Improving the basic model by the energy-flux delay

In the previous Sec. IV C we improved the classical source term accounting for the time-delay by the delay function $F_{\rm del}(t)$. Substituting the new form (29) of $\eta^{\rm cl}$ in the basic model Eq. (25), we formulate the "improved decay model" of quantum turbulence:

$$\frac{d\mathcal{L}}{dt} = \frac{2\,(d_2\Delta)^2 F_{\rm del}(t)}{\kappa^2(t+\tau_{\rm del}-\tau_2)^3} - \frac{\alpha\,\kappa}{d_1}\mathcal{L}^2\ . \quad (30)$$

For $\tau_{\rm dec} = 0$ the improved model (30) coincides with the basic model (25). For $t < \tau_{\rm del}$ the delay function (29b) $F_{\rm del}(t) < 1$ and the energy flux term in Eq. (30), that is proportional to $F(t)$, is suppressed. For $t \sim \tau_{\rm del}$ this term is gradually switching on and, finally, for $t \gg \tau_{\rm del}$ the improved and basic models, Eqs. (30) and (25) again coincide.

By analogy with Eq. (28), we can formulate the analytical form of the improved decay model:

$$\tilde{\mathcal{L}}(t) = f_{\rm del}(t)\mathcal{L}^{\rm cl}(t+\tau_{\rm del})\coth\left[\frac{f_{\rm del}(t)\mathcal{L}^{\rm cl}(0)}{\mathcal{L}^{\rm Q}}\right]\ . \quad (31)$$

Now we are fully armed to compare the suggested analytical models (28) and (31) with experimental observations. This comparison is the subject of the two following Sections.

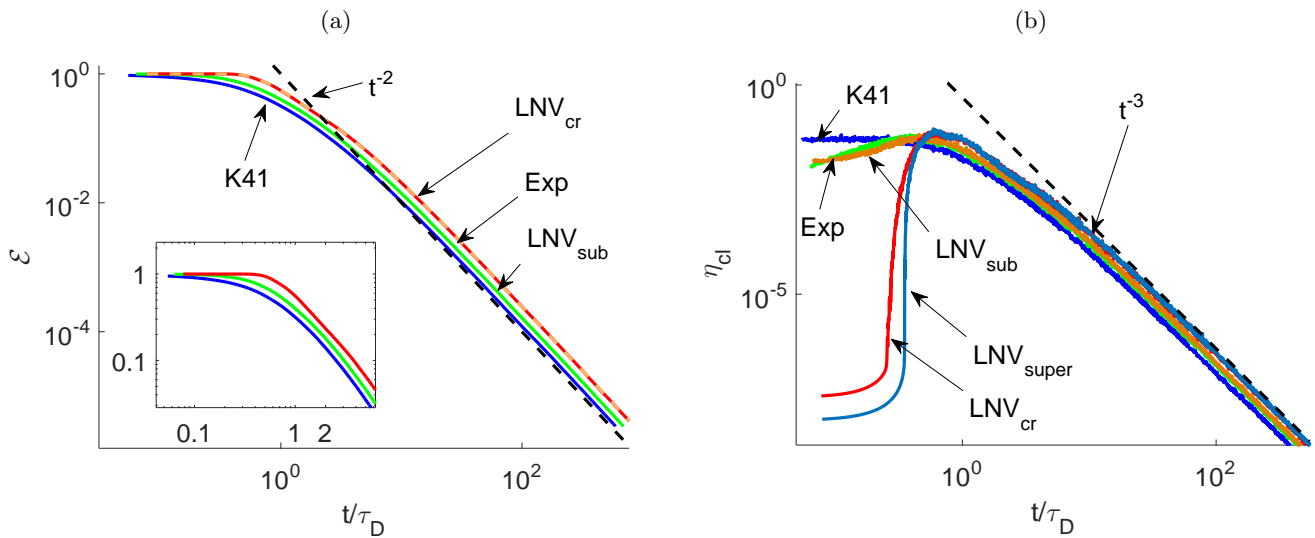

FIG. 8: (color online) Panel (a) – time dependence of the total energy $\mathcal{E}(t)$ in decaying superfluid turbulence from different initial conditions, shown in Fig. 7. The inset shows the short-time evolution of the total energy $\mathcal{E}(t)$. Panel (b) – time dependence of the rate of energy dissipation $\varepsilon(t)/\kappa^2$ that plays role of the VLD flux $\eta^{\rm cl}(t)$ in the model Eq. (20). The lines are colored as in Fig. 7 without markers. In panel(a), the lines for critical and subcritical LNV-spectra coincide. The lines for experimental and supercritical LNV-spectrum almost coincide.

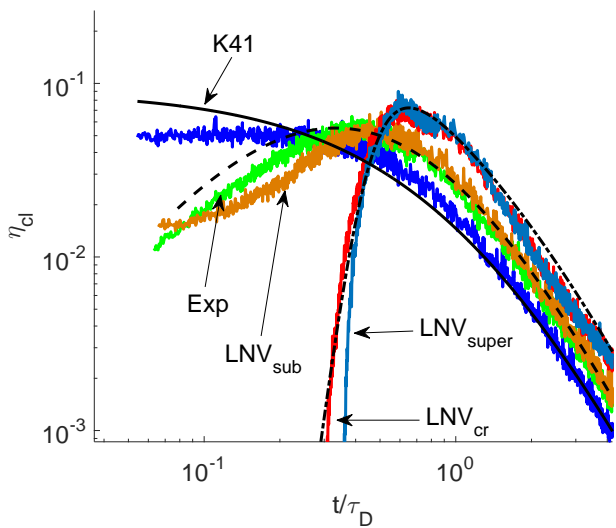

FIG. 9: (color online) The details of the short-time dependence of the rate of energy dissipation $\varepsilon(t)/\kappa^2$, shown in Fig. 8(b). Black solid line – the basic model dependence (24e) for $\eta^{\rm cl}(t)$ fits K41 initial conditions ICa. Black dashed line shows improved model dependence (29) with $n = 2$; it approximates numerically found $\eta^{\rm cl}(t)$ for weakly localized initial conditions ICb and ICe. Black dot-dashed line, given by (29) with $n = 6$, fits $\eta^{\rm cl}(t)$ computed with strongly localized initial conditions ICc and ICd.

#### 2. Basic decay model vs coflow experiment

We begin here with the more simple coflow case, choosing for comparison the less noisy set # 3, demonstrating (red line) in Fig. 2(a) the decay of almost 4 decades over 50 s. This line is reproduced (in red) in Fig. 10(a) together with the plot of analytical $\bar{\mathcal{L}}(t)$ (shown by the blue dotted line), predicted by the basic model. This line is barely seen because it practically coincides with the red experimental line within the line-width, broadened after 25 s by noise. Some discrepancy between the experiment and the model prediction is better seen on the closeup Fig. 10(b) showing the first 5 s of the decay.

Notice that the basic model is very simple: it completely ignores any interaction between the quantum peak and the classical large-scale turbulence. In formulating the model we also ignored the energy stored in the equilibrium part of the classical energy $\mathcal{E}_{\rm s}^{\rm TD}$. Last but not least, the model does not account for the spatial inhomogeneity of the turbulent channel flow, in which turbulent kinetic energy significantly depends on the distance to the walls. Bearing all these in mind, we consider the agreement between the experiment and the model quite satisfactory.

#### 3. Improved decay model vs counterflow experiment

Our next step is a discussion of the decay in the counterflow and pure superflow cases. For our rather simplified manner of modeling, there is no physical difference between the counterflow and pure superflow decaying turbulence. Therefore from the data presented in

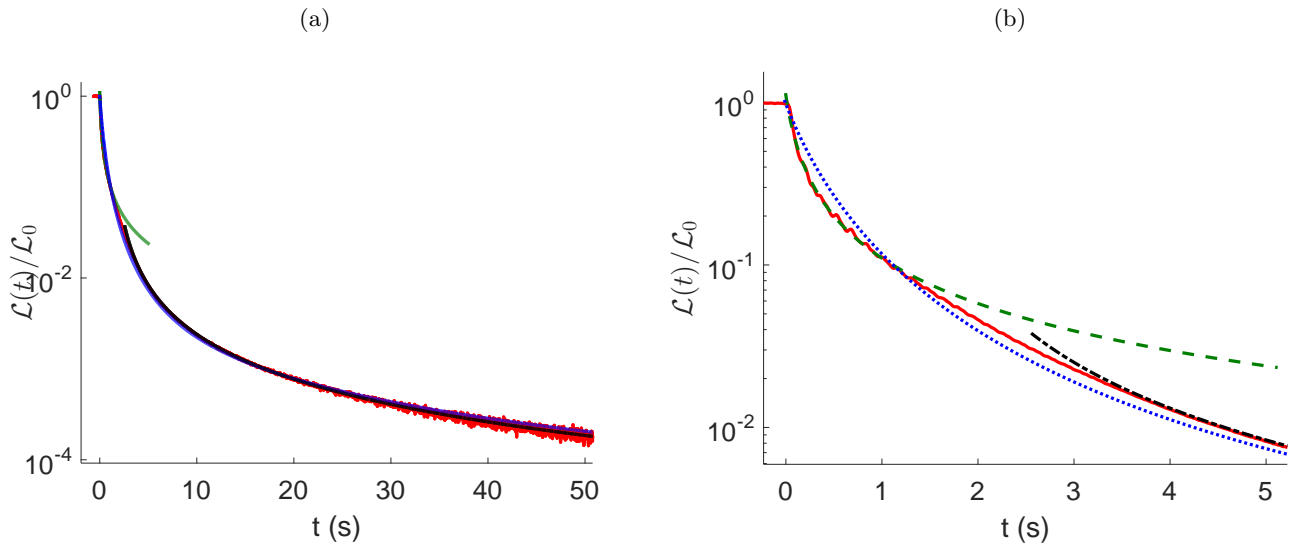

FIG. 10: (color online) Comparison of the experimental observation of coflow decay with the basic model predictions. Red solid lines, reproduced from red lines in Figs. 2(a,b), show the coflow data for set # 3 ($T = 1.35$ K, $\mathcal{L}_0 \approx 10^6$ cm$^{-2}$). Blue dotted lines show basic model prediction, $\bar{\mathcal{L}}(t)$, Eq. (28) with $\tau_1 = 0.5$ s, $\tau_2 = 1$ s and $b_2 = 0.075$.

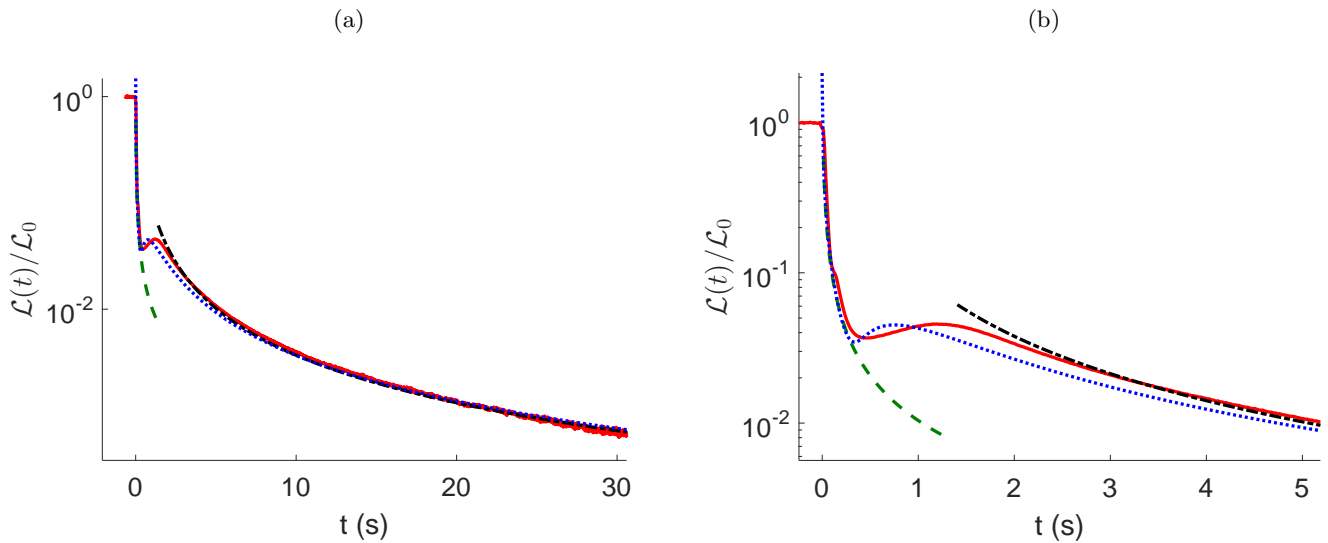

FIG. 11: (color online) Comparison of the experimental observation for the superflow with the improved model predictions. Red solid lines, reproduced from red lines in Figs. 3(c,d) for superflow # 12 ($T = 1.45$ K, $\mathcal{L}_0 \approx 10^6$ cm$^{-2}$). Blue dotted lines show improved model prediction, $\tilde{\mathcal{L}}(t)$, Eq. (31) with $n = 2$, $\tau_1 = 0.015$ s, $\tau_2 = 0.1$ s, $\tau_{\rm dec} = 0.6$ s and $b_2 = 0.22$.

Figs. 3, we chose the set # 12 (superflow at $T = 1.45$ K, $\mathcal{L}_0 \simeq 10^6$ cm$^{-2}$) shown in Figs. 3(c,d) by red lines. In this case the noise is relatively low and the bump on the $\mathcal{L}(t)$-dependence is clearly pronounced. Red lines in Figs. 11(c,d) reproduce the experimental results shown in Figs. 3(c,d) in red. Blue dotted lines result from the analytical form $\tilde{\mathcal{L}}(t)$ of the improved model of decay, given by Eq. (31) with $n = 2$, $\tau_1 = 0.015$ s, $\tau_2 = 0.1$ s, $b_2 = 0.22$.

Semi-quantitative agreement between experimental observation and the improved decay model allows us to conclude that this model reflects the basic physical mechanisms responsible for the time dependence of the VLD. In particular, it accounts for the time delay in the delivery of the energy flux from classical to quantum parts of superfluid turbulence.

## V. TOWARDS THEORY OF QUANTUM TURBULENCE WITH COUNTERFLOW

In mechanically driven quantum turbulence, the mean velocities of the normal and superfluid components are known to coincide: $\boldsymbol{U}_{\rm n} = \boldsymbol{U}_{\rm s}$. Numerous laboratory, numerical and analytical studies showed that under these conditions the mutual friction between the normal and superfluid velocity components couple also their fluctuations: $\bm{u}_{\rm n}(\bm{r},t) \approx \bm{u}_{\rm s}(\bm{r},t)$ almost at all scales. This is not the case in thermally driven quantum turbulence, where the counterflow velocity $\bm{U}_{\rm ns} \neq 0$ partially decouples the normal and superfluid velocity fluctuations and enhances the turbulent energy dissipation due to the mutual friction. In this Section we suggest a simple analytical model of the resulting energy balance in counterflow turbulence that predicts dramatic suppression of the energy spectrum at intermediate and small scales.

All ingredients of this model are already known. We just have to adopt Eqs. (13) for the one-dimensional energy spectra $\mathcal{E}_{\rm s}, \mathcal{E}_{\rm n}$ and to take in this equation expression for $\mathcal{E}_{\rm ns}(k)$ from Eq. (12). In this way we formulated the *Differential Model* for the normal and superfluid energy spectra $\mathcal{E}_{\rm n}(k)$ and $\mathcal{E}_{\rm s}(k)$ of counterflow turbulence, given by the closed set of differential Eqs. (13), and (12). These equations may be solved numerically.

### A. Energy spectra in "symmetric" counterflow turbulence with $\rho_{\rm n} = \rho_{\rm s}$

In order to get qualitative information on the form of the energy spectra $\mathcal{E}_{\rm n}(k)$ and $\mathcal{E}_{\rm s}(k)$, we consider the derived equations in the important limiting case $\Omega_{\rm ns} \gg \gamma_{\rm ns}, \nu_{\rm n} k^2, \nu'_{\rm s} k^2$. Then the steady-state version of Eqs. (13) and (12) simplify and take the form:

$$\frac{5}{8}\frac{d}{dk} k^{5/2} \mathcal{E}_{\rm s,n}^{3/2}(k) \qquad (32)$$
$$= \Omega_{\rm s,n} \Big\{ \frac{\Omega\, \mathcal{E}_{\rm s}(k) + \Omega_{\rm n}\, \mathcal{E}_{\rm n}(k)}{\Omega_{\rm ns}} D(k) - \mathcal{E}_{\rm s,n} \Big\}\,.$$

Dramatic further simplification occurs for $T \simeq 1.95\,{\rm K}$, when $\rho_{\rm n} \simeq \rho_{\rm s}$. In this case we expect $\mathcal{E}_{\rm s}(k) = \mathcal{E}_{\rm n}(k) \equiv \mathcal{E}$ and Eqs. (32) become an ordinary differential equation for $\mathcal{E}(k)$:

$$\frac{5}{8}\frac{d}{dk} k^{5/2} \mathcal{E}^{3/2}(k) = \Omega\, \mathcal{E}(k) \left[ D(k) - 1 \right], \qquad (33a)$$
$$\zeta_q = q/q_\times\,, \quad q_\times = 2\,\Omega/(k_0 V_{\rm ns})\,. \qquad (33b)$$

Here the dimensionless wave number $q = k/k_0$ with $k_0 \simeq \pi/\Delta$ and the outer scale of turbulence $\Delta$ is defined such that Eq. (32) are valid for $q > 1$. Introducing new dimensionless function

$$\Psi(q) \equiv \frac{15\, k_0^{3/2}}{8\,\Omega} \sqrt{q^{5/3} \mathcal{E}(q\, k_0)}\,, \qquad (34a)$$

$$\mathcal{E}(k) = \Big[\frac{8 \Psi(k/k_0)}{15}\Big]^2 \frac{\Omega^2}{k_0^{4/3} k^{5/3}}\,, \qquad (34b)$$

we get from Eqs. (33a) the equation

$$q^{5/3} \frac{d\,\Psi(q)}{dq} = D(q/q_\times) - 1\,, \qquad (34c)$$

which can be solved analytically with the boundary condition at $q = 1$ (i.e. $k = k_0$):

$$\Psi(1) = \Psi_0 = \frac{8\Omega}{15\sqrt{k_0^3 \mathcal{E}_0}}\,, \qquad (35a)$$

$$\Psi(q) = \Psi_0 + \frac{1}{q_\times^{2/3}} \Big[ I\Big(\frac{1}{q_\times}\Big) - I\Big(\frac{q}{q_\times}\Big) \Big], \qquad (35b)$$

$$I(z) = \int_0^z \frac{1 - D(y)}{y^{5/3}} dy\,. \qquad (35c)$$

Using now Eqs. (34b) and (35) one gets:

$$\mathcal{E}(k) = \mathcal{E}_0 \Big\{ 1 + A \Big[ I\Big(\frac{k}{k_\times}\Big) - I\Big(\frac{k_0}{k_\times}\Big) \Big] \Big\}^2 \Big(\frac{k_0}{k}\Big)^{5/3},\quad (36{\rm a})$$

$$A = \frac{8\,\Omega\, k_0^{2/3}}{15\, k_\times^{2/3} \sqrt{k_0^3 \mathcal{E}_0}}\,. \qquad (36b)$$

Integral (35c) can be found analytically:

$$I(z) = \frac{\sqrt{3}\pi}{5} + \frac{3}{20} \Big\{ -\frac{4}{z^{2/3}} - 2\sqrt{3}\arctan[\sqrt{3} - 2z^{1/3}]$$
$$-2\sqrt{3}\arctan[\sqrt{3} + 2z^{1/3}] + \frac{4\arctan[z]}{z^{5/3}}$$
$$+ \ln\Big[1 - \frac{3z^{2/3}}{(1+z^{2/3})^2}\Big] \Big\}\,. \qquad (36c)$$

It has the following asymptotics:

$$I(z) = \frac{z^{4/3}}{4}\Big(1 - \frac{6z^2}{25}\Big), \quad \text{for } z \ll 1\,, \qquad (37a)$$

$$I(z) = \frac{\sqrt{3}\,\pi}{5} - \frac{3}{2z^{2/3}}, \quad \text{for } z \gg 1\,. \qquad (37b)$$

Simple analytical expression

$$I(z) \simeq \frac{z^{4/3}}{4 + 0.7z^2}\,, \qquad (37c)$$

approximates the exact Eq. (36c), with relative accuracy within 3% in the $z < 2$ region, while expression

$$I(z) = \frac{\sqrt{3}\,\pi}{5} - \frac{3}{2z^{2/3}} + \frac{\pi}{4z^{5/3}}\,, \qquad (37d)$$

works well for $z > 2$.

Notice that for $k_0 < k \ll k_\times$ the energy spectrum (36) deviates down from the K41 $\frac{5}{3}$-spectrum:

$$\mathcal{E}(k) \simeq \mathcal{E}_0 \Big\{ 1 + \frac{A}{4\, k_\times^{4/3}} \big[k_0^{4/3} - k^{4/3}\big] \Big\}^2 \Big(\frac{k_0}{k}\Big)^{5/3}\,. \quad (38{\rm a})$$

The spectrum for $k \gg k_\times$ crucially depends on the value of $A$. There exists a critical value

$$A_{\rm cr} = 1/I(\infty) = 5/(\pi\sqrt{3}) \approx 0.92\,. \qquad (38b)$$

For $A < A_{\rm cr}$ the system asymptotically tends to K41 spectrum

$$\mathcal{E}(k) \simeq \mathcal{E}_0 \Big\{1 - \frac{A}{A_{\rm cr}}\Big\}^2 \Big(\frac{k_0}{k}\Big)^{5/3}, \qquad (38c)$$

but with the energy flux $\varepsilon_\infty = \varepsilon_0(1 - A/A_{\rm cr})^3$, smaller than the energy input rate $\varepsilon_0$. The difference $(\varepsilon_0 - \varepsilon_\infty)$ is dissipated by mutual friction. This is similar to the subcritical LNV spectrum (18) of $^3$He turbulence with resting normal fluid component.

For $A > A_{\rm cr}$, $\mathcal{E}(k) = 0$ for large $k$. In the differential approximation used here, the spectrum $E(k)$ sharply terminates at some finite $k_*$, in the same manner as the supercritical LNV $^3$He spectrum (19):

$$\mathcal{E}(k) \propto \frac{1}{k^{5/3}} \Big\{\frac{1}{k^{2/3}} - \frac{1}{k_*^{2/3}}\Big\}^2. \qquad (38d)$$

The cutoff wave number $k_*$ may be found from the equation

$$1 = A[I(\zeta_{k_*}) - I(\zeta_0)] \approx A\, I(\zeta_{k_*}). \qquad (38e)$$

When $A \to A_{\rm cr}$, $k_* \to \infty$ and $\mathcal{E}(k) \propto k^{-3}$ at large $k$, exactly like in the critical LNV $^3$He spectrum (17).

The energy spectra (36) for different values of $A$ in the range $0 \leqslant A \leqslant 6$, that includes the critical value $A_{\rm cr} \simeq 0.92$, are shown in Fig. 12a. We see that increase in the mutual friction force, characterized in Eq. (36) by the dimensionless parameter $A$, suppresses the energy spectra from the Kolmogorov-41 behavior $\mathcal{E}(k) \propto k^{-5/3}$ (for $A=0$) towards the critical spectrum $\mathcal{E}(k) \propto k^{-3}$ at $A = A_{\rm cr} \simeq 1$. Further increase in $A$ localizes energy spectra in the $k$-space, as shown in Fig. 3(a). There is clear qualitative similarity of the energy spectra in conterflow turbulence with the LNV spectra. This allowed us to use the LNV spectra in Sec. IV C for the analysis of the delay function $F_{\rm del}(t)$ in our improved model of the VLD decay.

### B. Velocity structure functions in "symmetric" counterflow turbulence

Recent visualization experiment in counterflow[37] reported that transversal velocity structure function $S_{2,\perp}(r) \propto r$ in an interval about one decade, adjacent to the outer scale of turbulence. To see how these observations may be rationalized using our model, we consider the second order velocity structure functions, defined as follows:

$$S_2(\boldsymbol{r}) = \langle |\boldsymbol{u}(\boldsymbol{r}) - \boldsymbol{u}(0)|^2 \rangle, \qquad (39a)$$
$$S_{2,\perp}(\boldsymbol{r}) = \langle [u_\perp(\boldsymbol{r}) - u_\perp(0)]^2 \rangle, \qquad (39b)$$
$$S_{2,||}(\boldsymbol{r}) = \langle [u_{||}(\boldsymbol{r}) - u_{||}(0)]^2 \rangle. \qquad (39c)$$

Here $u_\perp$ and $u_{||}$ are projections of the turbulent velocity $\boldsymbol{u}$ on directions orthogonal and parallel to the separation $\boldsymbol{r}$. In isotropic turbulence, assumed above in the analytical model of the spectra, all structure functions depend only on $r = |\boldsymbol{r}|$ and all are proportional to each other. Up to a numerical factor of the order unity, they may be expressed via one-dimensional energy spectrum $\mathcal{E}(k)$ as follows:

$$S_2(r) \simeq \int_0^\infty \mathcal{E}(k)[1 - \sin(kr)/(kr)]\, dk. \qquad (40a)$$

With the Kolmogorov-41 spectrum $\mathcal{E}_{\rm K41}(k) = \mathcal{E}_0(k_0/k)^{5/3}$ [Eq. (36) with $A = 0$], it gives the classical result

$$S_2(r) \simeq 1.2 V_{\rm T}^2 (k_0 r)^{2/3}, \quad V_{\rm T} \equiv \sqrt{k_0 \mathcal{E}_0}, \qquad (40b)$$

shown by the upper green dashed straight line in Fig. 12c. Next, we account for the fact that in reality the available range of $k$ is limited: $k_{\min} < k < k_{\max}$. The value of $k_{\min} \sim \pi/\Delta$ is determined by the outer scale of turbulence. For quantum turbulence in the superfluid component $k_{\max} \simeq \pi/\ell$. In the Florida experiments[37], our estimates show that $k_{\min} \simeq 0.3 k_\times$ and $k_{\max} \simeq 200 k_\times$. Replacing limits in the integral (40a) by these values, we compute again $S_2(r)$ with the K41 spectrum- see upper green solid line in Fig. 12c. We observe the same scaling behavior $S_2(r) \propto r^{2/3}$ in the interval of about two decades (from $r \simeq 0.05$ to $r \simeq 5$). For nonzero values of $A$ the log-log plots of $S_2(r)$ versus $r$ can be considered as approximately straight lines with the slope that increases with $A$. In particular, for $A = 0.6$ $S_2(r)$, shown in Fig. 12c by solid red line, it is practically indistinguishable from the straight line with the slope $+1$ (shown by dashed red line) in the interval $0.02 < r < 2$. This means that, for $A = 0.6$, $S_2(r) \propto r$ with high accuracy in the interval of two decades. To see this better, in Fig. 12d we present the plot of $S_2(r)$ compensated by $1/r$. The solid red line in Fig. 12(d) lays indeed very close to the black thin horizontal line.

Notice that the energy spectrum for $A = 0.6$, used to find $S_2(r)$ (red solid line in Fig. 12b), is essentially different from the scale-invariant spectrum $\mathcal{E}(k) \propto k^{-2}$, shown in Fig. 12b by the dashed blue line. Using spectrum $\mathcal{E}(k) \propto k^{-2}$, we computed $S_2(r)$ again, with the result shown in Fig. 12d by the dashed blue line. Unexpectedly, this result demonstrates scale-invariant behavior $S_2(r) \propto r$ on a shorter range. We also computed $S_2(r)$ using supercritical LNV spectrum (18), with $k_{\rm cr} = 1.15 k_0$. This spectrum, shown in Fig. 12b by the green dash-dotted line, is very different from the $1/k^2$ behavior (blue dashed line). Nevertheless, the resulting structure function $S_2(r)$ (green dash-dotted line in Fig. 12d) again demonstrates the scale-invariant behavior of $S_2(r) \propto r$ over more than two decades.

We conclude that very different energy spectra, including the spectrum (36) with $A \simeq 0.6$, found here, can result in the reported Ref.[37] $S_2(r) \propto r$ with somewhat smaller extent of the scaling behavior, of about one decade. This means that our analytical model does not contradict the observation[37]. Nevertheless, at this stage we are not in the position to claim that the model explains the observed simple behavior $S_2(r) \propto r$. We will return to this point in conclusive Sec. VI.

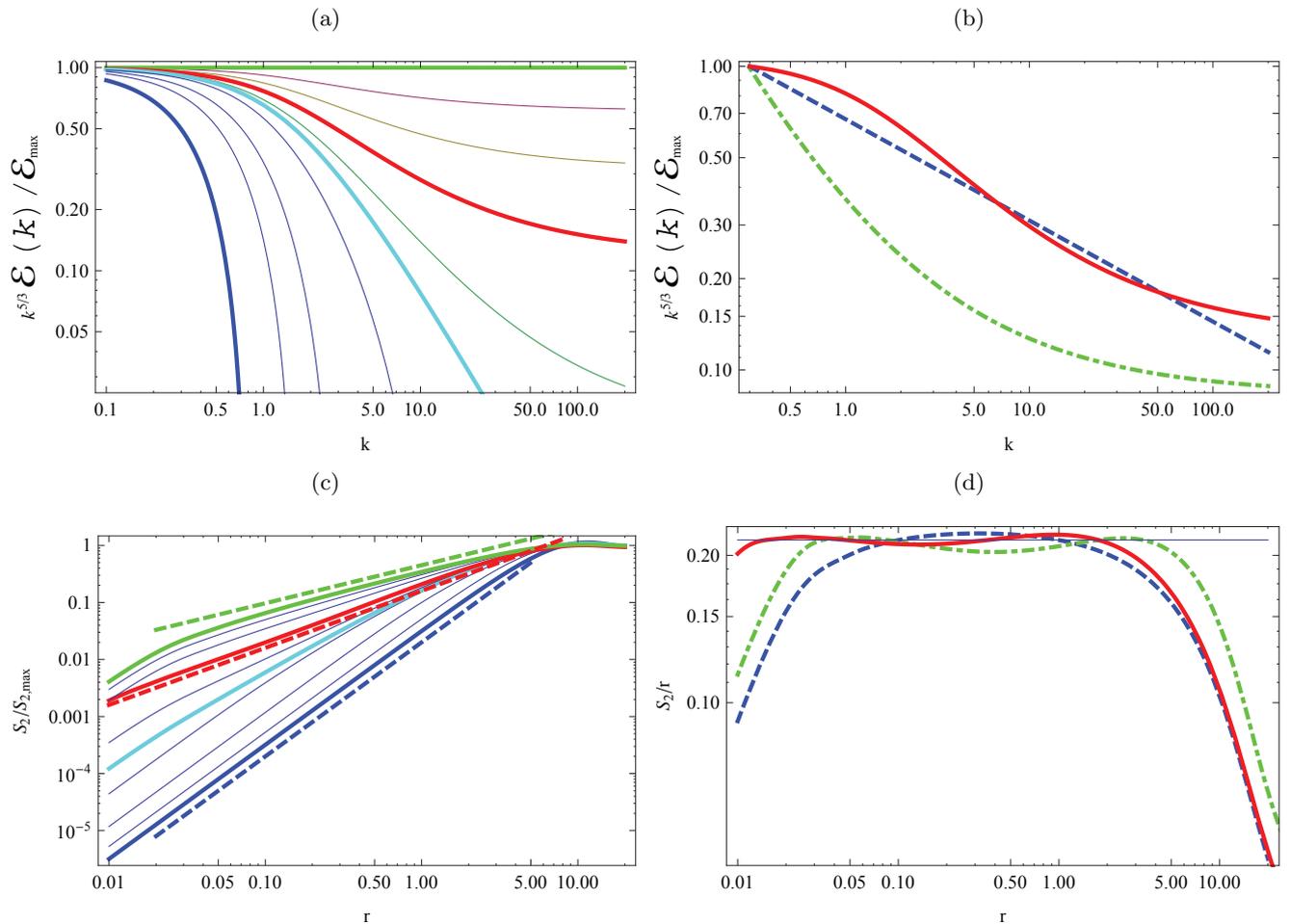

FIG. 12: Color online. Panel (a) Log-Log plots of compensated normalized energy spectra $k^{5/3}\mathcal{E}(k)/\mathcal{E}_{\max}$ for $k_0 = 0$ and with different values of $A$, starting with $A = 0$ (upper green line), through $A = 0.2, 0.4, 0.6$ (red thick line) $0.8, 0.92$ (thick cyan line – critical value), $1, 1.5, 2.0\,3.0$ and last line – $A = 6$. Panel (b) Log-Log plots of compensated normalized energy spectrum $k^{5/3}\mathcal{E}(k)/\mathcal{E}_{\max}$ for $k_0 = 0.3 k_\times$ with $A = 0.6$ (solid red line), $1/k^2$ (after compensation by $k^{5/3}$) – blue dashed line and LNV subcritical spectrum $k^{-3}[1 + (k/k_*)^{2/3}]^2$ with $k_* = 1.15 k_\times$ – green dash-dotted line. Panel (c) Log-log plots of the second order structure functions $S_2(r)$ (normalized by their large $r$ limit), computed with Eq. (40a) and with the same values of $A$ [and the same color code] as in Panel (a). The thick dashed straight lines indicate scaling laws: upper green line $\propto r^{2/3}$ ), middle red line $\propto r$ and lower blue line $\propto r^2$). Panel(d) The normalized compensated by $1/r$ structure functions $S_2(r)$, calculated using the spectra shown in Panel(b) (with the same color code).

Here we notice only that, as the parameter $A$ increases, the energy spectra become more and more localized at small $k$, as seen in Fig. 12(a), while the apparent slope of the corresponding structure functions increases, tending to 2. This is clearly seen in Fig. 12(c). The $S_2(r)$ for the largest $A = 6$ is shown by solid blue line in comparison with the scaling function $r^2$, shown by the blue dashed line. The reason for such a behavior is simple: for large $A$, the energy spectrum terminates at small $k$ and for scales smaller than $1/k$, the velocity field becomes smooth and differentiable. Thus the velocity field can be expanded in the Taylor series, the velocity difference across the separation $r$ is proportional to $r$, and the structure function is proportional to $r^2$.

### C. Approximations of the simple analytical model

It should be stressed that in our approach we adopted some uncontrolled approximations and simplifications, widely used in the studies of classical hydrodynamic turbulence. Among them are celebrated hypotheses, suggested by Kolmogorov in 1941, concerning the small scale turbulent statistics:

44

1. universality (independence of the energy pumping);
2. isotropy;
3. locality of energy transfer over scale.

These hypotheses have been justified in numerous experiments, numerical and analytical studies of developed turbulence of classical fluids, mechanically driven at large scales. However, very little is known about turbulent statistics in the case of thermally driven quantum counterflow turbulence.

We can quite easily accept the first hypothesis of the universality, for instance in the wide and long enough channels with reasonably well-controlled surface of the wall.

The second assumption is the isotropy hypothesis. It is known from numerical simulations of (mainly space homogeneous) counterflow turbulence that the anisotropy of the quantized vortex tangle is rather small (about 10%), see e.g. Ref. [38,41,43,44]. However very little is known about anisotropy of the velocity fluctuations on scales $R$ above the intervortex distance $\ell$. Without this knowledge and having clear understanding that the assumption of isotropy on scales $R > \ell$ may be questionable, we nevertheless assume isotropy of turbulence statistics from the very beginning. This simplifying assumption allows us to formulate a simple analytical model in terms of one-dimensional energy spectrum $\mathcal{E}(k)$, which is an angular average of the full (and possibly anisotropic) three-dimensional energy spectrum.

The third assumption is the locality of the energy transfer, which is build-in the algebraic closure (14b), used in our approach. Bearing in mind that the nonlinear terms in the two-fluid equations of motion for the normal and superfluid velocity components are the same as in the Navier-Stokes equation, and, moreover, if the resulting energy spectrum is not very different from the classical $\frac{5}{3}$-scaling for the classical fluid, we can accept also this assumption. The detailed analysis of the problem of locality in classical turbulence with the scale-invariant scaling $\mathcal{E}(k) \propto k^{-x}$ shows [45,46] that in the "window of locality" $3 < x < 5$ the energy transfer is really local. For our case this means that for the subcritical spectra (with $A < 1$, when the the local slope $x(k) = d\ln[\mathcal{E}(k)]/d\ln k$ is within the window of locality), this assumption is reasonable. We also hope that the strong version of the locality assumption, the algebraic closure (14b), leads to qualitatively correct energy spectra. The situation with the supercritical spectra is less simple. Definitely, the consequence of Eqs. (36) and (38d) that $\mathcal{E}(k) = 0$ for $k > k_*$, is an artefact of the algebraic closure. We think that for large $k$, the supercritical energy spectra will instead decay very fast with $k$, presumably $\propto k^{-y}$ with $y > 5$, as a result of the direct (non-local) energy transfer to the $k$-region from the intermediate region of scales about $k_{\rm int}$, where the local slope $x(k_{\rm int}) \simeq 5$.

Last but not least, an additional restriction of our approach is full ignorance of possible long-living coherent structures at scales $\gtrsim \ell$ that may contribute to the statistics of quantum turbulence with counterflow. There is a great deal of speculation in the studies of classical space-homogeneous hydrodynamic turbulence, but, nevertheless, the question of their statistical relevance is still open. Moreover, the well justified multi-fractal models of classical turbulence (see, e.g., textbook [47]) describe in many details the statistics of classical turbulence without any direct reference to coherent structures. Coherent structures in quantum turbulence can be characterized simply as "terra incognita".

In such a situation, much more experimental, numerical and analytical work is required to formulate a theory of quantum turbulence which will account for the interplay of coexisting classical and quantum forms of superfluid turbulent energy in all relevant details. Nevertheless, we consider our experimental findings and simple analytical models of steady-state and decaying quantum turbulence as a natural and perhaps even required step in a long way toward desired level of understanding and description of the basic physical mechanisms that govern quantum turbulence.

## VI. SUMMARY

Being motivated by the challenge to understand the quantum turbulence occurring in superfluid $^4$He, we report this complementary experimental, numerical and theoretical treatise of turbulent coflow, counterflow and pure superflow of superfluid $^4$He in a channel. The level of agreement between the experimental observations and the analytical predictions for the time evolution of the vortex-line density in decaying turbulence, demonstrated in this paper, allows us to conclude that the developed basic and improved models adequately reflect the underlying physical processes responsible for the decay of quantum turbulence, originating from various types of steady superfluid $^4$He flows (coflow, counterflow, pure superflow), including
– the interplay of classical and quantum processes, resulting in two decay laws of VLD;
– the partial decoupling of the normal and superfluid velocity fields in pure superflow and counterflow turbulence, predicted in Ref. [23];
– the resulting suppression of energy spectra in these flows leading to the time-delay in the energy flux from classical to quantum length scales of turbulence in superfluid $^4$He.

Being inspired by these findings, we made the first step towards the theory of steady-state, space homogeneous turbulence of counterflowing superfluid $^4$He. The suggested by us simple analytical theory results in the energy spectra $\mathcal{E}(k, A)$ given by Eq. (36) and shown in Fig. 12a. These spectra depend on the dimensionless parameter $A$, that describes the intensity of the mutual friction, responsible for the coupling of the normal and superfluid velocity, relative to the counterflow velocity, responsible for their decoupling. For the particular value $A \simeq 0.6$, the energy spectrum $\mathcal{E}(k, 0.6)$, shown in Fig. 12b is close to $1/k^2$ scaling and results in the second-order structure function $S_2(r) \propto r$ over the interval of about two decades, in agreement with the observations reported in Ref. [37].

# APPENDIX A: SABRA-SHELL MODEL OF QUANTUM TURBULENCE

A detailed study of the decay of large-scale turbulence has been done in the framework of the so-called Sabra-shell model of quantum turbulence [31,34,42]:

$$\left[\frac{d}{dt} + \nu_{\rm n} k_m^2\right] u_m^{\rm n} = {\rm NL}[u_m^{\rm n}] + F_m^{\rm n}, \quad ({\rm A1a})$$

$$\left[\frac{d}{dt} + \nu_{\rm s} k_m^2\right] u_m^{\rm s} = {\rm NL}[u_m^{\rm s}] - F_m^{\rm s}, \quad ({\rm A1b})$$

$${\rm NL}[u_m] = i\Big(a k_{m+1} u_{m+2} u_{m+1}^* \quad ({\rm A1c})$$

$$+ b k_m u_{m+1} u_{m-1}^* - c k_{m-1} u_{m-1} u_{m-2}\Big),$$

$$F_m^{\rm s} = \Omega(u_m^{\rm s} - u_m^{\rm n}) \quad ({\rm A1d})$$
$$F_m^{\rm n} = \Omega_{\rm n}(u_m^{\rm s} - u_m^{\rm n}).$$

These equations represent a simplified version of the coarse-grained, two-fluid, gradually-damped Hall-Vinen-Bekarevich-Khalatnikov (HVBK) equations in the $\boldsymbol{k}$-representation. They mimic the statistical behavior of $\boldsymbol{k}$-Fourier components of the turbulent superfluid and normal velocity fields in the entire shell of wave vectors $k_m < k < k_{m+1}$ by complex shell velocity $u_m^{\rm n,s}$. The shell wave numbers are chosen as a geometric progression $k_m = k_0 \lambda^m$, where $m = 1, 2, \dots M$ are the shell indexes, and we have used the shell-spacing parameter $\lambda = 2$, $k_0 = 1/16$ and $M = 28$ shells.

Similarly to the HVBK (and the Navier-Stokes or Euler equation), the ${\rm NL}[u_m]$ term in Eq. (A1c) is quadratic in velocities, proportional to $k$ and conserves (in the forceless, inviscid limit) the kinetic energy $E = \frac{1}{2}\sum_m |u_m|^2$, provided that $a + b + c = 0$. We used here the Sabra version[48] of ${\rm NL}[u_m]$ with the traditional (and physically motivated) choice $b = c = -a/2$, which describes important features of superfluid turbulence including intermittency corrections [31].

In Eq. (A1a), $\nu_{\rm n} = \mu/\rho_{\rm n}$ is the kinematic viscosity of the normal component, i.e., its dynamical viscosity $\mu$, normalized by the normal fluid density. The effective superfluid viscosity $\nu_{\rm s}$ describes the energy sink in superfluids, e.g. due to the vortex reconnections. For more details of the origin and role of $\nu_{\rm s}$ see Ref. [2,31]. The mutual friction terms $F_m^{\rm n,s}$, given by Eq (A1d), is just the $k$-Fourier transform of Eqs. (10a) and (10b).

As a first step in our study of the large-scale turbulence decay we can simplify the procedure further, using the fact that the turnover time in our situations is longer than the coupling time $1/\Omega_{\rm ns}$ given by Eq. (10d). Therefore the first stage of the decay, during which the normal- and superfluid velocities become coupled, is short with respect to the time required for the developing of the Richardson-Kolmogorov cascade. Skipping this stage, we can assume that the velocities are fully coupled according to Eq. (8). In this case Eqs. (A1) turn into one Sabra-equation for $u_m = u_m^{\rm s} = u_m^{\rm n}$:

$$\left[\frac{d}{dt} + \nu k_m^2\right] u_m = {\rm NL}[u_m], \quad ({\rm A2})$$

with $\nu = (\mu + \nu_{\rm s} \rho_{\rm s})/\rho$.

The equations (A2) were solved using the adaptive time step 4th order Runge-Kutta with exponential time differencing[49]. For more details see Ref. [34,42]. The evolution of the system was followed for about $10^3 \tau_\Delta$. In all cases we perform $10^4$ simulations with the same initial total energy, but different, randomly distributed phases of initial shell-velocities and perform ensemble averaging over initial conditions.

# APPENDIX B: SOME DEFINITIONS AND KNOWN RELATIONSHIPS

To define the one-dimensional energy spectra $\mathcal{E}_{\rm n}(k)$, $\mathcal{E}_{\rm s}(k)$ and cross-correlation $\mathcal{E}_{\rm ns}(k)$ we need to recall some definitions and relationships, that are well-known in statistical physics. The first is the set of Fourier transforms in the following normalization:

$$\boldsymbol{u}'_{\rm n,s}(\boldsymbol{r}, t) \equiv \int \frac{d\boldsymbol{k}}{(2\pi)^3} \boldsymbol{v}_{\rm n,s}(\boldsymbol{k}, t) \exp(i \boldsymbol{k} \cdot \boldsymbol{r}), \quad ({\rm B1a})$$

$$\boldsymbol{v}_{\rm n,s}(\boldsymbol{k}, t) \equiv \int \frac{d\omega}{2\pi} \widetilde{\boldsymbol{v}}_{\rm n,s}(\boldsymbol{k}, \omega) \exp(-i\omega t), \quad ({\rm B1b})$$

$$\widetilde{\boldsymbol{v}}_{\rm n,s}(\boldsymbol{k}, \omega) = \int d\boldsymbol{r} dt\, \boldsymbol{u}'_{\rm n,s}(\boldsymbol{r}, t) \exp[i(\omega t - \boldsymbol{k} \cdot \boldsymbol{r})]. \quad ({\rm B1c})$$

The same normalization is used for other objects of interest.

Next we define the simultaneous correlations and cross-correlations in $\boldsymbol{k}$-representation, (proportional to $\delta(\boldsymbol{k} - \boldsymbol{k}')$ due to homogeneity):

$$\langle \boldsymbol{v}_{\rm n}(\boldsymbol{k}, t) \cdot \boldsymbol{v}_{\rm n}^*(\boldsymbol{k}', t) \rangle = (2\pi)^3 E_{\rm nn}(\boldsymbol{k}) \delta(\boldsymbol{k} - \boldsymbol{k}'), \quad ({\rm B2a})$$

$$\langle \boldsymbol{v}_{\rm s}(\boldsymbol{k}, t) \cdot \boldsymbol{v}_{\rm s}^*(\boldsymbol{k}', t) \rangle = (2\pi)^3 E_{\rm ss}(\boldsymbol{k}) \delta(\boldsymbol{k} - \boldsymbol{k}'), \quad ({\rm B2b})$$

$$\langle \boldsymbol{v}_{\rm n}(\boldsymbol{k}, t) \cdot \boldsymbol{v}_{\rm s}^*(\boldsymbol{k}', t) \rangle = (2\pi)^3 E_{\rm ns}(\boldsymbol{k}) \delta(\boldsymbol{k} - \boldsymbol{k}'). \quad ({\rm B2c})$$

We also need to define cross-correlations $\langle \widetilde{\boldsymbol{v}}_{\rm n} \cdot \widetilde{\boldsymbol{v}}_{\rm s}^* \rangle$ in $(\boldsymbol{k}, \omega)$-representation:

$$\langle \widetilde{\boldsymbol{v}}_{\rm n}(\boldsymbol{k}, \omega) \cdot \widetilde{\boldsymbol{v}}_{\rm s}^*(\boldsymbol{k}', \omega') \rangle \quad ({\rm B3a})$$
$$= (2\pi)^4 \widetilde{E}_{\rm ns}(\boldsymbol{k}, \omega) \delta(\boldsymbol{k} - \boldsymbol{k}') \delta(\omega - \omega').$$

This object is related to the simultaneous $\langle \boldsymbol{v}_{\rm n} \cdot \boldsymbol{v}_{\rm s}^* \rangle$ cross-correlation (B2c) via the frequency integral:

$$\langle \boldsymbol{v}_{\rm n}(\boldsymbol{k}, t) \cdot \boldsymbol{v}_{\rm s}^*(\boldsymbol{k}', t) \rangle = \int d\omega \widetilde{E}_{\rm ns}(\boldsymbol{k}, \omega). \quad ({\rm B3b})$$

Here and below "tilde" marks the objects defined in $(\boldsymbol{k}, \omega)$-representation.

It is known also that the $\boldsymbol{k}$-integration of the correlations (B2) produces their one-point second moment:

$$\int \frac{d\boldsymbol{k}}{(2\pi)^3} E_{\rm nn}(\boldsymbol{k},t) = \left\langle |\boldsymbol{u}_{\rm n}(\boldsymbol{r},t)|^2 \right\rangle, \qquad (B4a)$$

$$\int \frac{d\boldsymbol{k}}{(2\pi)^3} E_{\rm ss}(\boldsymbol{k},t) = \left\langle |\boldsymbol{u}_{\rm s}(\boldsymbol{r},t)|^2 \right\rangle, \qquad (B4b)$$

$$\int \frac{d\boldsymbol{k}}{(2\pi)^3} E_{\rm ns}(\boldsymbol{k},t) = \left\langle \boldsymbol{u}_{\rm n}(\boldsymbol{r},t) \cdot \boldsymbol{u}_{\rm s}(\boldsymbol{r},t) \right\rangle. \qquad (B4c)$$

In the isotropic case, each of the three correlations $E_{\dots}(\boldsymbol{k})$ is independent of the direction of $\boldsymbol{k}$: $E_{\dots}(\boldsymbol{k}) = E_{\dots}(k)$ and $\int \dots d\boldsymbol{k} = 4\pi \int \dots k^2\, dk$. This allows the introduction of the one-dimensional energy spectra $\mathcal{E}_{\rm s}$, $\mathcal{E}_{\rm n}$ and the cross-correlation $\mathcal{E}_{\rm ns}$ as follows:

$$\mathcal{E}_{\rm n}(k,t) = \frac{k^2}{2\pi^2} E_{\rm nn}(k,t), \quad \mathcal{E}_{\rm s}(k,t) = \frac{k^2}{2\pi^2} E_{\rm ss}(k,t),$$
$$\mathcal{E}_{\rm ns}(k) \equiv \frac{k^2}{2\pi^2} E_{\rm ns}(k,t). \qquad (B5)$$

# ACKNOWLEDGEMENTS

We thank W.F. Vinen and Wei Guo for interesting discussions. This work is supported by the Czech Science Foundation under project GAČR 203/14/02005S and by the European Community Framework Programme 7, EuHIT - European High-performance Infrastructures in Turbulence, grant agreement no. 312778.

[1] C. F. Barenghi, L. Skrbek, and K. R. Sreenivasan, Proc. Natl. Acad. Sci. USA **111**, 4647 (2014).
[2] W. F. Vinen and J. J. Niemela, J. Low Temp. Phys. **128**, 167 (2002).
[3] L. Skrbek and K.R. Sreenivasan, Phys. Fluids **24**, 011301 (2012).
[4] R. J. Donnelly, *Quantized Vortices in Hellium II*, (Cambridge University Press, Cambridge, 1991).
[5] *Quantized Vortex Dynamics and Superfluid Turbulence*, edited by C. F. Barenghi, R. J. Donnelly and W .F. Vinen, Lecture Notes in Physics **571**, (Springer-Verlag, Berlin, 2001).
[6] L. Skrbek and K. R. Sreenivasan, in *Ten Chapters in Turbulence*, edited by P. A. Davidson, Y. Kaneda, and K. R. Sreenivasan, (Cambridge University Press, Cambridge, 2013), pp. 405437.
[7] C. F. Barenghi, V. S. L'vov, and P.-E. Roche, Proc. Natl. Acad. Sci. USA **111**, 4683 (2014).
[8] P. L. Walstrom, J. G. Weisend, J. R. Maddocks, and S. W. Van Sciver, Cryogenics **28**, 101 (1988).
[9] S. Babuin, E. Varga, and L. Skrbek, J. Low Temp. Phys. **175**, 324 (2014).
[10] S. Babuin, E. Varga, L. Skrbek, E. Leveque, and P.-E. Roche, Europhys. Lett. **106**, 24006 (2014).
[11] E. Varga, S. Babuin, and L. Skrbek, Phys. Fluids **27**, 065101 (2015).
[12] M. R. Smith, R. J. Donnelly, N. Goldenfeld, and W. F. Vinen, Phys. Rev. Lett. **71**, 2583 (1993).
[13] S. R. Stalp, L. Skrbek, and R. J. Donnelly, Phys. Rev. Lett. **82**, 4831 (1999).
[14] H. E. Hall and W. F. Vinen, Proc. Roy. Soc. A **238**, 204 (1956).
[15] K. W. Schwarz and J. R. Rozen, Phys. Rev. Lett. **66**, 1896 (1991).
[16] H. Tennekes and J. L. Lumley, *A first course in turbulence*, (The MIT Press, 1994).
[17] H. Schlichting and K. Gersten, *Boundary-layer theory*, (Springer, 2000).
[18] M. L. Baehr and J. T. Tough, Phys. Rev. Lett. **53**, 1669 (1984).
[19] S. S. Courts and J. T. Tough, Phys. Rev. B **38**, 74 (1988).
[20] W. F. Vinen, Proc. Roy. Soc. A **240** 114, 128 (1957); **242** 489, 493 (1957).
[21] S. Babuin, E. Varga, W. F. Vinen, and L. Skrbek, Phys. Rev. B **92**, 184503 (2015).
[22] J. Gao, W. Guo, V. S. Lvov, A. Pomyalov, L. Skrbek, E. Varga, and W. F. Vinen, JETP Letters, **103**, 732 (2016).
[23] D. Khomenko, V. S. L'vov, A. Pomyalov, and I. Procaccia, Phys. Rev. B, **93** 014516 (2016).
[24] S. Babuin, M. Stammeier, E. Varga, M. Rotter, and L. Skrbek, Phys. Rev. B **86**, 134515 (2012).
[25] L. Skrbek, A. V. Gordeev, and F. Soukup, Phys. Rev. E **67**, 047302 (2003).
[26] A. V. Gordeev, T. V. Chagovets, F. Soukup and L. Skrbek, J. Low Temp. Phys. **138**, 549 (2005).
[27] R. J. Donnelly and C. F. Barenghi, J. Phys. Chem. Ref. Data **27**, 1217 (1998).
[28] S. B. Pope, *Turbulent Flows,* (Cambridge University Press, Cambridge, England (2000).
[29] W. F. Vinen, Phys. Rev. B, **61** 1410 (2000).
[30] V. S. L'vov, S. V. Nazarenko, and L. Skrbek, J. Low Temp. Phys. **145**, 125 (2006).
[31] L. Boue, V. S. L'vov, Y. Nagar, S. V. Nazarenko, A. Pomyalov and I. Procaccia, Phys. Rev. B **91**, 144501 (2015).
[32] C. Connaughton and S. Nazarenko, Phys. Rev. Lett. **92**, 044501 (2004).
[33] V. S. L'vov, S. V. Nazarenko, and G. E. Volovik, JETP Letters **80**, 479 (2004).
[34] L. Boue, V. S. L'vov, A. Pomyalov, and I. Procaccia, Phys. Rev. B **85**, 104502 (2012).
[35] L. Kovasznay, J. Aeronaut. Sci. **15**, 745 (1947).
[36] W. F. Vinen, Phys. Rev. B **71**, 024513 (2005).
[37] A. Marakov, J. Gao, W. Guo, S. W. Van Sciver, G. G.

Ihas, D. N. McKinsey, and W. F. Vinen, Phys. Rev B **91**, 094503 (2015).
[38] K. W. Schwarz, Phys. Rev. B **38**, 2398 (1988).
[39] L. Skrbek, A. V. Gordeev, and F. Soukup, Phys. Rev. E **67**, 047302 (2003).
[40] D. Khomenko, V.S. L'Vov, A. Pomyalov, and I. Procaccia, Physical Review B **93**, 134504(2016)
[41] L. Kondaurova, V. S. L'vov, A. Pomyalov, and I. Procaccia, Phys. Rev. B, **89**, 014502 (2014).
[42] L. Boué, V. S. L'vov, A. Pomyalov, and I. Procaccia, Phys. Rev. Lett. **110**, 014502 (2013).
[43] R. T. Wang, C. E. Swanson, and R. J. Donnelly, Physical Review B **36**, 5240 (1987).
[44] H. Adachi, S. Fujiyama, and M. Tsubota, Phys.Rev. **81**, 10451 (2010).
[45] V.S. L'vov and G. E. Falkovich, Phys. Rev. A 46 (8) 4762-4772 (1992).
[46] V.S. L'vov and I. Procaccia. Phys. Rev. E, **52**, 3840 (1995)
[47] U. Frisch, *Turbulence: The Legacy of A. N. Kolmogorov*, (Cambridge University Press, 1995).
[48] V. S. L'vov, E. Podivilov, A. Pomyalov, I. Procaccia, and D. Vandembroucq, Phys. Rev. E **58**, 1811 (1998).
[49] S. M. Cox and P. C. Matthews, J. Comput. Phys. **176**, 430 (2002).